\documentclass[useAMS,usenatbib,usegraphicx]{mn2e}

\usepackage{graphicx,times,epsfig,rotating,amsmath}

\def\mnras{MNRAS}
\def\aj{AJ}
\def\aap{A\&A}
\def\apj{ApJ}
\def\apjl{ApJ}
\def\apjs{ApJS}
\def\araa{ARA\&A}
\def\pasp{PASP}
\def\nat{Nature}

\def\aaps{A\&AS}

\newcommand{\kmsend}{\mbox{km s$^{-1}$}}
 \newcommand{\kms}{\mbox{km s$^{-1}$ }}

\newcommand{\hpc}{\mbox{h$^{-1}$pc }}

\newcommand{\msun}{\mbox{M$_{\sun}$ }}
\newcommand{\msunend}{\mbox{M$_{\sun}$}}

\newcommand{\cmthree}{\mbox{cm$^{-3}$}}

\newcommand{\msunyr}{\mbox{M$_{\sun}$yr$^{-1}$ }}
\newcommand{\msunyrend}{\mbox{M$_{\sun}$yr$^{-1}$}}
\newcommand{\htwo}{\mbox{H$_2$}}

\newcommand{\z}{\mbox{$z$}}

\newcommand{\zsim}{\mbox{$z\sim$ }}

\newcommand{\magorrian}{\mbox{$M_{\rm BH}$-$M_{\rm bulge}$ }}
\newcommand{\lbol}{L$_{\rm {bol}}$}

\newcommand{\sunrise}{\mbox{\sc sunrise}}
\newcommand{\gadget}{\mbox{\sc gadget-3}}
\newcommand{\gadgettwo}{\mbox{\sc gadget-2}}
\newcommand{\starburst}{\mbox{\sc starburst99}}
\newcommand{\mappings}{\mbox{\sc mappingsiii}}

\newcommand{\bzk}{\mbox{{\it BzK}}}

\newcommand{\nhtwo}{\mbox{$N_{\rm H2}$}}
\newcommand{\xco}{\mbox{$X_{\rm CO}$}}
\newcommand{\xcounits}{\mbox{cm$^{-2}$/K-km s$^{-1}$}}

\newcommand{\alphaco}{\mbox{$\alpha_{\rm CO}$}}
\newcommand{\alphacounits}{\mbox{\msun pc$^{-2}$ (K-\kmsend)$^{-1}$}}

\title[The Effect of Galactic Environment on \xco]{A
  General Model for the CO-\htwo \ Conversion Factor in Galaxies
   with Applications to the Star Formation Law}

\author[Narayanan et al.]{Desika\, Narayanan$^{1}$\thanks{E-mail:
    dnarayanan@as.arizona.edu}\thanks{Bart J. Bok Fellow}, Mark R.
  Krumholz$^{2}$, Eve C. Ostriker$^3$, Lars Hernquist$^4$\\$^{1}$Steward
  Observatory, University of Arizona, 933 N Cherry Ave, Tucson, Az,
  85721\\$^{2}$Department of Astronomy and Astrophysics, University
  of California, Santa Cruz, Ca, 95064\\$^{3}$Department of
  Astronomy, University of Maryland, College Park, Md
  20742\\$^{4}$Harvard-Smithsonian Center for Astrophysics, 60 Garden
  St., Cambridge, Ma 02138\\}
\begin{document}

\date{Accepted by MNRAS}

\pagerange{\pageref{firstpage}--\pageref{lastpage}} \pubyear{2010}

\maketitle

\label{firstpage}

\begin{abstract}

The most common means of converting an observed CO line intensity into
a molecular gas mass requires the use of a conversion factor (\xco).
While in the Milky Way this quantity does not appear to vary
significantly, there is good reason to believe that \xco \ will depend
on the larger-scale galactic environment.  With sensitive instruments
pushing detections to increasingly high redshift, characterising \xco
\ as a function of physical conditions is crucial to our understanding
of galaxy evolution.  Utilising numerical models, we investigate how
varying metallicities, gas temperatures and velocity dispersions in
galaxies impact the way CO line emission traces the underlying \htwo
\ gas mass, and under what circumstances \xco \ may differ from the
Galactic mean value.  We find that, due to the combined effects of
increased gas temperature and velocity dispersion, \xco \ is depressed
below the Galactic mean in high surface density environments such as
ULIRGs.  In contrast, in low metallicity environments, \xco \ tends to
be higher than in the Milky Way, due to photodissociation of CO in
metal-poor clouds.  At higher redshifts, gas-rich discs may have
gravitationally unstable clumps that are warm (due to increased star
formation) and have elevated velocity dispersions.  These discs tend
to have \xco \ values ranging between present-epoch gas-rich mergers
and quiescent discs at low-\z.  This model shows that {\it on
  average}, mergers do have lower \xco \ values than disc galaxies,
though there is significant overlap.  \xco \ varies smoothly with the
local conditions within a galaxy, and is not a function of global
galaxy morphology. We combine our results to provide a general fitting
formula for \xco \ as a function of CO line intensity and metallicity.
We show that replacing the traditional approach of using one constant
\xco \ for starbursts and another for discs with our best-fit function
produces star formation laws that are continuous rather than bimodal,
and that have significantly reduced scatter.

\end{abstract}
\begin{keywords}
ISM:clouds-ISM:molecules-galaxies:interactions-galaxies:ISM-galaxies:starburst-galaxies:star
formation

\end{keywords}

\section{Introduction}
\label{section:introduction}

As the building block of stars, \htwo \ is arguably the most important
molecule in astrophysics.  Ironically, however, it is also one of the
more observationally elusive.  With no permanent dipole moment, \htwo
\ is best directly detected via its first quadrupole line.  This line
lies at $\sim500 $ K above ground, significantly above the $\sim10$ K
typical of the cold molecular interstellar medium (ISM), and in a
spectral region with relatively low atmospheric transmission.
As a result, giant molecular clouds (GMCs) are often studied via
tracer molecules. The ground-state rotational transition of carbon
monoxide ($^{12}$CO (J=1-0), hereafter CO) is one of the most common
tracers of \htwo \ in GMCs owing to its relatively high abundance
\citep[$\sim 10^{-4}/$\htwo \ in the Galaxy;][]{lee96}, the high
atmospheric transmission at $\sim3$ mm where the J=1-0 line lies, and
the low temperatures and densities required for CO excitation ($\sim
5$ K, $\sim 10^2-10^3$ \cmthree).  However, using CO to trace \htwo
\ does not come without uncertainty.  At the basis of the
interpretation of CO observations is the conversion between CO
spectral line intensity and \htwo \ gas mass, the so-called CO-\htwo
\ conversion factor.

The CO-\htwo \ conversion factor is defined as
\begin{equation}
\xco = \frac{\nhtwo}{W_{\rm CO}}
\label{equation:xco}
\end{equation}
where \nhtwo \ is the molecular gas column density, and $W_{\rm CO}$
is the velocity-integrated CO line intensity (measured in K-\kmsend).
Alternatively, the conversion factor can be defined as the ratio of
the molecular gas mass and CO line luminosity:
\begin{equation}
\alpha_{\rm CO} = \frac{M_{\rm gas}}{L_{\rm CO}}
\end{equation}
\xco \ and \alphaco \  are easily related via 
\begin{equation}
\begin{split}
\xco \ ({\rm cm}^{-2} (\rm K-\kmsend)^{-1})= \\
6.3 \times 10^{19} \times \alpha_{\rm CO} (\msun {\rm pc}^{-2}({\rm K}-\kmsend)^{-1})
\end{split}
\end{equation}
Hereafter, we refer to the CO-\htwo \ conversion factor in terms of
\xco\footnote{In the literature, \xco \ is sometimes referred to as
  the $X$-factor, and we shall use the two interchangeably.}, though
plot in terms of both \xco \ and $\alpha_{\rm CO}$.  

Despite potential variations in CO abundances, radiative transfer
effects, and varying \htwo \ gas fractions, a variety of independent
measurements of \htwo \ gas mass in GMCs have shown that the CO-\htwo
\ conversion factor in Galactic clouds is reasonably constant, with
$\xco \approx 2-4 \times 10^{20}$ \xcounits \ ($\alphaco \approx 3-6 \ 
\alphacounits$).  Methods for obtaining independent
measurements of \htwo \ gas mass include (i) assuming the GMCs are
in virial equilibrium, and utilising the CO line width to derive the
\htwo \ mass within the CO emitting region
\citep[e.g. ][]{lar81,sol87}; (ii) Inferred dust masses and an assumed
dust-to-gas mass ratio \citep{dic75, dev87, gue93,
  dam01,dra07,lom06,pin08,ler11,mag11} and (iii) $\gamma$-ray emission
arising from the interaction of cosmic rays with \htwo
\ \citep{blo86,ber93,str96,hun97,abd10,del11}.  Beyond this,
observations of GMCs in the Local Group suggest that a similar
CO-\htwo \ conversion factor may apply for some clouds outside of our
own Galaxy \citep{ros03,bli07,don11}. Numerical models of molecular
clouds on both resolved and galaxy-wide scales have indicated that a
relatively constant CO-\htwo \ conversion factor in the Galaxy and
nearby galaxies may naturally arise from GMCs that have a limited
range in surface densities, metallicities and velocity dispersions
\citep[][]{glo11,she11a,nar11b,fel11b}.

In recent years, a number of observational studies have provided
evidence for at least two physical regimes where the CO-\htwo \
conversion factor departs from the ``standard'' Milky Way value.  The
first is in high-surface density environments.  Interferometric
observations of present-epoch galaxy mergers by
\citet{sco91,dow93,sol97,dow03,hin06,mei10} and \citet{dow98} showed
that using a Milky Way \xco \ would cause the inferred \htwo \ gas
mass to exceed the dynamical mass of the CO-emitting region for some
galaxies.  This implies that the CO-\htwo \ conversion factor should
be lower than the Galactic mean in high-surface density environments.
More recent observations of \zsim2 Submillimetre Galaxies (SMGs) by
\citet{tac08} suggested a similar result for high-redshift starbursts.
Similarly, observations of GMCs toward the Galactic Centre indicate
that \xco \ may be lower in this high-surface density environment
\citep{oka98}.

Second, in low-metallicity environments at both low and high-\z, the
CO-\htwo \ conversion factor may be larger than the Milky Way mean
value \citep{wil95,ari96,isr97,bos02,ler06,bol08,ler11,gen11b}, though
there is some debate over this \citep[see summaries in ][and the
  Appendix of \citet{tac08}]{bli07}.  Observations have suggested that
the $X$-factor may scale as $X \propto$ (O/H)$^{-b}$ where
$b = 1-2.7$ \citep{ari96,isr97}.

The fact that these two effects drive \xco \ in opposite directions
complicates the interpretation of CO detections from high-redshift
systems where galaxies display a large range in metallicities
\citep[e.g. ][]{sha04,gen11b,sha11} and gas surface densities
\citep[e.g. ][]{bot09,dad10b,gen10,nar11a}.  Further muddying the
interpretation of high-\z \ molecular line emission is the fact that
there are not always clear analogs of high-redshift galaxies in the
present-day Universe.  For example, relatively unperturbed discs at
\zsim 2 oftentimes have surface densities, star formation rates, and
velocity dispersions comparable to local galaxy mergers
\citep{dad05,dad10b,kru10,gen11a}, though (sometimes) lower
metallicities \citep{cre10}.  Similarly, even the most heavily
star-forming galaxies at \zsim 2, SMGs, at times show dynamically cold
molecular discs even when they are potentially the result of mergers
\citep{nar09,car10,nar10a,eng10}.  Converting CO line intensity to
\htwo \ gas masses is a multi-faceted problem that involves
understanding how galactic environment sets the $X$-factor.

 Over the last two decades, models of GMC evolution have made
 substantive headway in elucidating the variation of \xco \ with
 physical properties of molecular clouds.  The earliest GMC models
 utilised 1D radiative transfer with spherical cloud models \citep[e.g
 ][]{kut85,wal07}.  Photodissociation region (PDR) models furthered
 these studies by including the formation and destruction pathways of
 CO \citep{bel06,bel07,mei07}. More recently, magnetohydrodynamic
 models of GMC evolution with time-dependent chemistry by
 \citet{glo10} and \citet{glo11} coupled with radiative transfer
 calculations \citep{she11a,she11b} have investigated \xco \ on the
 scales of individual GMCs, and its dependence on the physical
 environment.

Compared to models of isolated GMCs, there are relatively few
simulations exploring the effect of the larger galactic environment on
the $X$-factor.  \citet{mal88} presented some of the earliest models
which explored the effects of changing individual physical parameters
in isolation on the CO-\htwo \ conversion factor.  Very recently,
\citet{fel11b} have tied the GMC models of \citet{glo11}
to cosmological simulations of galaxy evolution to investigate \xco
\ on galaxy-wide scales in relatively quiescent disc galaxies.
Building on these models, as well as what has been learned in the
studies of \citet{glo11}, \citet{she11a} and \citet{wol10}, in
\citet{nar11b}, we combined dust and molecular line radiative transfer
calculations with hydrodynamic simulations of galaxies in evolution in
order to develop a model that aims to capture the CO line emission
from GMCs on galaxy-wide scales.

In this paper, utilising the methodology we developed in
\citet{nar11b}, we investigate the effect of galactic environment in
setting the CO-\htwo \ conversion factor in galaxies at low and
high-redshifts.  Our paper is organised as follows.  In
\S~\ref{section:methods}, we summarise the methodology developed in
\citet{nar11b} and employed here.  We note that while
\S~\ref{section:methods} is a shorter summary, a more complete
description is presented in Appendix~\ref{section:appendix}.  In
\S~\ref{section:environment}, we investigate the role of galactic
environment on \xco, focusing on isolated disc galaxies
(\S~\ref{section:isolated}), galaxies at low metallicity
(\S~\ref{section:metallicity}), high-surface density
(\S~\ref{section:surfacedensity}), and high redshift
(\S~\ref{section:highredshift}).  Building on these results, in
\S~\ref{section:observations}, we develop a functional form for
calculating \xco \ from observations of galaxies and, as an
application, utilise these to interpret Kennicutt-Schmidt star
formation rate relations at low and high-redshift.  In
\S~\ref{section:discussion}, we discuss our results in the context of
other theoretical models, and in \S~\ref{section:summary}, we
summarise our main results.

\section{Summary of Simulation Methodology}
\label{section:methods}

Our main goal is to simulate the impact of galactic environment on the
\htwo \ content and CO emission from galaxies.  This involves
simulating the evolution of galaxies, the physical state of the
molecular ISM, and the radiative transfer of CO lines through GMCs and
through galaxies.  Because much of the methodology has been described
in a previous paper by us \citep{nar11b}, we briefly summarise our
approach here, and defer the quantitative details to the Appendix.

We first require model galaxies to analyse.  We simulate the
hydrodynamic evolution of disc galaxies in isolation and galaxy
mergers over a range of galaxy masses, merger mass ratios, and
redshifts utilising a modified version of the publicly available code,
\gadgettwo.  These simulations provide information regarding the
kinematic structure, mass and metal distribution of the ISM, as well
as the stellar populations.  Table A1 summarises the model properties. 

The remainder of our calculations occur in post-processing.  We smooth
the SPH results onto an adaptive mesh.  We require knowledge of the
physical and chemical state of the molecular clouds in our model
galaxies.  We assume the cold \htwo \ gas is bound in spherical GMCs with
\htwo \ fractions calculated following the models of \citet{kru08} and
\citet{kru09a}.  Carbon is assumed to have a uniform abundance within
these clouds of $1.5 \times 10^{-4} \times Z'$, where $Z'$ is the
metallicity with respect to solar.  The fraction of carbon locked up
in CO is determined following the models of \citet{wol10}, and have an
explicit dependence on the metallicity.  GMCs that have a surface
density greater than 100 \msun pc$^{-2}$ are considered resolved.
GMCs that are not resolved (typically low mass GMCs in large cells in
the adaptive mesh) have a floor surface density of the aforementioned
value imposed, consistent with observations of Local Group GMCs
\citep[e.g. ][]{bol08,fuk10}. Unresolved GMCs have velocity dispersions
equal to the virial velocity of the GMC, whereas resolved GMCs have
velocity dispersions determined directly from the simulations.


 The temperature of the GMCs is determined via a balance of the
 various heating processes on the gas (here, photoelectric effect,
 cosmic rays and energy exchange with dust), and line cooling. The
 dust temperature is calculated via the publicly available dust
 radiative transfer code, \sunrise \ \citep{jon10a}.  The cosmic ray
 heating rate is assumed take on the mean Galactic value (except as
 described in Appendix~\ref{section:cosmicrays}).  Utilising this model, the
 temperature ranges typically from $\sim 10$ K in quiescent GMCs to
 $>100 $ K in the centres of starbursts \citep{nar11b}.

Once the physical and chemical state of the ISM is known, we are
prepared to model the CO line emission via radiative transfer
calculations. The emergent CO line emission from the GMCs is
calculated via an escape probability formalism \citep{kru07}.  This
radiation is then followed through the galaxy in order to account for
radiative transfer processes on galaxy-wide scales
\citep{nar06b,nar11b}.  With these calculations, we know the thermal
and chemical state of the molecular ISM in our model galaxies, the
synthetic broadband SEDs, and the modeled CO emission line spectra.
At this point, we are in a position to understand variations in the
CO-\htwo \ conversion factor. We remind the reader that further
details regarding the implementation of these models can be found in
the Appendix.

We note that, in order to alleviate
confusion between simulation points and observational data on plots,
we will employ a system throughout this work in which filled symbols
exclusively refer to observational data, and open symbols refer to
simulation results.

\section{The Effect of of Galactic Environment on \xco}
\label{section:environment}

Our general goal is to understand how \xco \ depends on
the physical
environment in galaxies, and how it may vary with observable
properties of galaxies.  In order to do this, we must first develop
intuition as to how various physical parameters affect \xco.
In this section, we examine how \xco \ varies from the
Galactic mean in low-metallicity environments, high-surface density
environments, and at high-redshift.

Quantitatively, we define the mean \xco \ from a galaxy as:
\begin{equation}
\langle \xco \rangle  = \frac{\int \Sigma_{\rm H2} \ {\rm dA} }{\int W_{\rm CO} \ {\rm dA}}
\end{equation}
which is equivalent to a luminosity-weighted \xco \ over all GMCs in
the galaxy.  

\subsection{Review of \xco \ in \z=0 Quiescent Disc Galaxies}
\label{section:isolated}
We begin by considering \xco \ in disc galaxies at \z=0 with
metallicities around solar ($Z' \approx 1$).  These galaxies have mean
$X$-factors comparable to the Galactic mean and will serve as the
``control'' sample from which we will discuss variations in physical
parameters. 

Recalling \S~\ref{section:methods}, and referring to the Appendix,
when clouds are not resolved, we impose a floor surface density of 100
\msun pc$^{-2}$.  In these GMCs, the velocity dispersion is the virial
velocity of the cloud.  In our model \z=0 disc galaxies, gas
compressions within the galaxy are unable to cause significant
deviations from these subresolution values of $N_{\rm H2} \sim 10^{22}
$cm$^{-2}$ and $\sigma \sim 1-5 $ \kmsend. In this regime, the
temperatures of the GMCs typically fall to $\sim 8-10$ K.  This is the
usual temperature where cosmic rays dominate gas heating in our model;
the densities are not sufficiently high for any of the heating
processes to increase the gas temperatures drastically. At $\sim$solar
metallicities, a sufficient column of dust can easily build up to
protect the CO from photodissocation, and most of the carbon in
molecular clouds is in the form of CO.  With these modest conditions
in the clouds, and little variation throughout the galaxy \citep[see
  Figure 2 of ][]{nar11b}, the modeled \xco \ tends to be $\sim$ a few
$\times 10^{20}$ \xcounits \ (i.e. similar to the Milky Way mean),
with the only notable exception being GMCs toward the galactic centre
\citep{nar11b}.

As pointed out by \citet{nar11b}, while various subresolution
techniques are folded into our model disc galaxies, that we see
relatively little variation in the GMC properties in our model \z=0
discs is a statement that the galactic environment is not extreme
enough to cause significant deviations from the default surface
densities and velocity dispersions.  The temperatures and velocity
dispersions are allowed to vary freely with galactic environment, and
the surface densities have a floor value similar to actual GMCs
\citep{bli07,bol08}.  As shown by \citet{she11a}, GMCs with physical
parameters comparable to those observed in Galactic GMCs exhibit \xco
\ values close to the Galactic mean, $\xco  \approx 2-4 \times 10^{20}
\xcounits$ ($\alphaco \approx 3-6 \ \alphacounits$).  As we will show,
in mergers in the present Universe, low metallicity galaxies, and in
some cases, high-redshift discs, the physical properties of GMCs vary
sufficiently that this is no longer the case.

\subsection{The Effects of Metallicity on \xco}
\label{section:metallicity}

\begin{figure}
\hspace{-1.5cm}
\includegraphics[angle=90,scale=0.4]{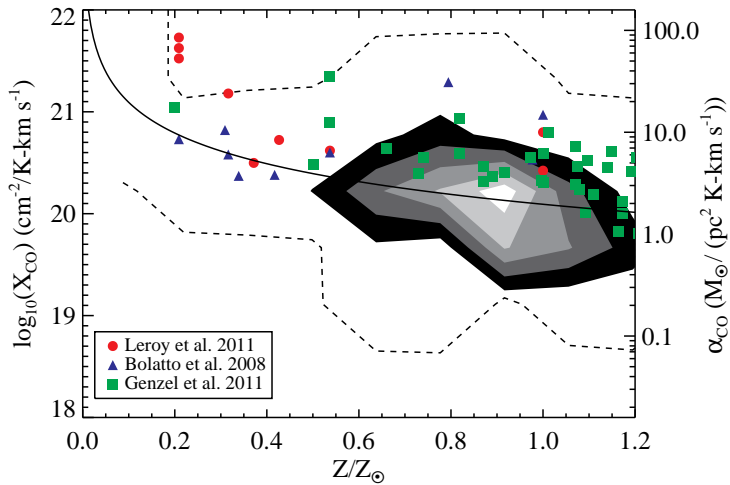}
\caption{\xco \ versus mass-weighted mean metallicity (in units of
  solar) for all \z=0 model galaxies with $\Sigma_{\rm H2} \sim 100$
  \msun pc$^{-2}$.  The contours represent the number of snapshots in
  a given \xco-$Z'$ bin, with the numbers increasing with increasing
  lightness of the contour.  The dashed line outer contour encompasses
  all model galaxies, regardless of their gas surface
  density. Overlaid are observational data points from \citet{bol08,
    ler11} and \citet{gen11b}.  The solid line shows our best fit to
  the simulations and is expressed in
  Equation~\ref{eq:xcozsd_ico_lazy} and described in
  \S~\ref{section:observations}.  \label{figure:xco_metallicity}}
\end{figure}

\begin{figure}
\hspace{-0.5 cm}
\includegraphics[angle=90,scale=0.4]{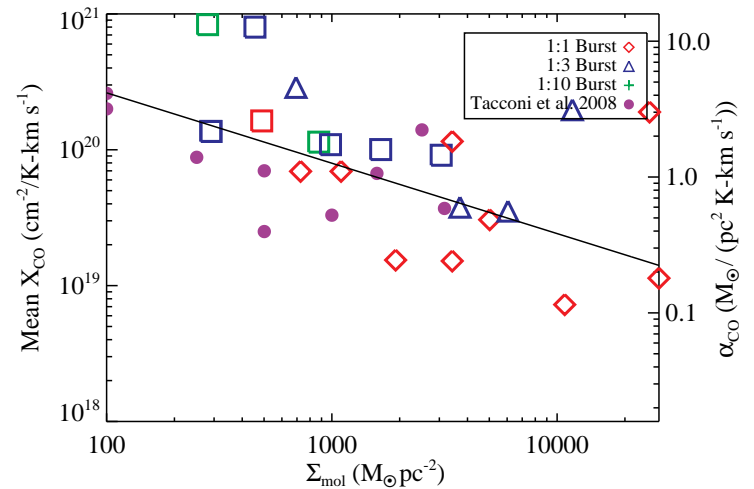}
\caption{Mean \xco \ versus mass-weighted mean surface density for
  \z=0 model mergers when they are undergoing a starburst (e.g. the
  snapshots with the peak SFRs).  The plotting symbols are shown in
  the legend, though a given symbol may be replaced by a square if the
  metallicity is lower than 0.5 $Z_\odot$.  The purple filled circles
  represent the compiled data from \citet{tac08}.  The black solid
  line is the best fit for all simulation snapshots from
  Equation~\ref{eq:xcozsd_ico_formal}, and is discussed in
  \S~\ref{section:observations}. \label{figure:xco_surfacedensity}}
\end{figure}

The metallicity of the gas can have a strong effect on the CO-\htwo
\ conversion factor.  In metal-poor gas, it is possible to have
``CO-dark'' molecular clouds \citep{pap02b,wol10}.  In these regions, \htwo
\ can self-shield to protect itself from photodissociating UV
radiation, whereas CO cannot and requires dust to survive
\citep{ste95,hol99}.  In these cases, we expect a larger fraction of
CO-dark clouds and a rise in the CO-\htwo \ conversion
factor \citep[][]{mal88,wol10,she11b,fel11b}.

In Figure~\ref{figure:xco_metallicity}, we plot the emission-weighted
mean \xco \ for our model \z=0 isolated disc galaxies as a function of
metallicity with filled contours. To control for surface density
effects, we plot models with $\Sigma_{\rm H2} \approx 100 \ \msun {\rm
}^{-2}$. The contours indicate the number of snapshots in a given
\xco$-Z'$ bin.  The outer dashed contour denotes the contour that
encompasses all of our models, at any gas surface density.

The lowest metallicity model galaxies all have relatively high mean
$X$-factors compared to the $\sim$solar metallicity galaxies.  The
lowest metallicity models can have mean $X$-factors approximately an
order of magnitude greater than the Galactic mean. As the galaxies
evolve and the ISM becomes enriched with metals, the carbon is almost
exclusively in the form of CO, and the $X$-factor decreases.

The predicted \xco-$Z'$ relation in
Figure~\ref{figure:xco_metallicity} matches well with recent
observational data.  Overlaid, we plot recent results from resolved
regions in nearby galaxies from \citet{bol08}, \citet{ler11} and
\citet{gen11b}.  Note that $\Sigma_{\rm H2}$ (and thus \xco) \ in the
observations depends on an estimate of the gas mass, which is obtained
with various methods for these studies\footnote{\citet{bol08} assume
  clouds are virialised; if instead clouds are marginally bound, then
  \xco \ would decrease.  \citet{ler11} use infrared emission to
  derive dust masses, and HI observations to derive a dust-to-gas
  ratio . If this dust to gas ratio does not map to the \htwo \ gas,
  the derived \xco \ will change accordingly.  \citet{gen10} assume a
  scaling relation $\Sigma_{\rm SFR} \propto \Sigma_{\rm gas}^{1.1}$.
  If a steeper relation were adopted (c.f. \S~\ref{section:kslaw}),
  then the values of \xco \ corresponding to high values of
  $\Sigma_{\rm SFR}$ would decrease somewhat.}.  Both the observations
and models show an upward trend in \xco \ with decreasing
metallicity.  While we defer a discussion of fitting \xco \ in terms
of $Z'$ to \S~\ref{section:observations}, we denote our best fit model
(discussed in Equation~\ref{eq:xcozsd_ico_lazy}) by the solid line in
Figure~\ref{figure:xco_metallicity}.

We should note that the contours in
Figure~\ref{figure:xco_metallicity} indicate the range of possible
$X$-factors at a given metallicity as returned by our models.  They
are not cosmological, so do not connote any particular probabilities.
The simple fact that observed data points lie within the contours with
a similar \xco-$Z'$ trend suggest a reasonable match between our
models and observed galaxies.  When examining the outer contour that
encompasses all of our models (and not just those at a given surface
density as in the greyscale contours), it is clear that there is a
significant dispersion in \xco \ on either side of the best-fit line
in Figure~\ref{figure:xco_metallicity}.  Moreover, there is a
significant dispersion in the observed data at a given metallicity
(especially noticeable near $Z'\approx 1$).  This suggests that there
is a second parameter controlling the $X$-factor in galaxies, aside
from metallicity.  As we will show in the next section, this is
the dynamical and thermal state of the molecular ISM.

\subsection{High Surface Density Galaxies}
\label{section:surfacedensity}

\subsubsection{Large Temperatures and Velocity Dispersions}
On average, in regions of high surface density, the emission-weighted
mean \xco \ in a galaxy is lower than the Galactic mean value of
$\sim2-4 \times 10^{20}$ \xcounits.  This has been shown
observationally by \citet{tac08}, and seen in the models of
\citet{nar11b}.

By itself, an increase in surface density does not cause \xco \ to
decrease.  Rather, the opposite is true: per
Equation~\ref{equation:xco}, at high surface densities, \xco \ would
{\it increase} if $W_{\rm CO}$ were fixed.  However, in
high-$\Sigma_{\rm H2}$ regions, the velocity-integrated line
intensity, $W_{\rm CO}$ in fact increases even more rapidly than
$\Sigma_{\rm H2}$, causing a net decrease in \xco.  To see why this
is, consider the physical processes that are typically associated
with increased surface densities.

First, regions of high surface density are associated with higher star
formation rates \citep{ken98a,kru09b,ost10,ost11}.  While a variety of
theories exist as to why this is the case \citep[see a review of some
  of these models in][]{tan10}, in our simulations this is due to the
fact that we explicitly tie our star formation rates to the volumetric
gas density on small scales.  With high star formation rates come
hotter dust temperatures as the UV radiation heats the nearby dust
grains. When the gas densities are $\ga 10^4$\cmthree, the dust and
gas exchange energy efficiently, and the gas temperature approaches
the dust temperature \citep{gol01,juv11}.  Hence, in regions of high
surface density, the increased dust temperature driven by the higher
star formation rates causes an increase in the gas kinetic
temperature.  Because the CO (J=1-0) line is thermalised at relatively
low densities, the brightness temperature of the line is comparable to
the kinetic temperature of the gas, and is thus increased in regions
of high surface density.  For low redshift galaxies, the easiest way
to increase the surface density tends to be through merging activity
and the associated tidal torques which drive gaseous inflows into the
nuclear region \citep{bar91,bar96,mih94a,mih96}, though
higher-redshift discs can have relatively large surface densities
simply due to gravitational instabilities in extremely gas-rich clumps
\citep{spr05a,bou10}.

Second, in the simulations, high surface-densities in gas are
typically accompanied by high velocity dispersions.  This generally
means some level of merging activity in low-redshift galaxies, and
either merging activity or unstable gas clumps in high-redshift discs.
In major mergers, the emission-weighted velocity dispersions can be as
high as $\sim 50-100 $ \kmsend.

The increased velocity dispersion and kinetic temperature contribute
roughly equally to increasing the velocity-integrated line intensity.
During a merger, the temperature and velocity dispersion individually
increase enough to offset the increase in the gas surface density, and
in combination tend to drive the emission-weighted mean \xco \ for a
galaxy below the Milky-Way average.  In regions of high surface
density, the emission-weighted mean \xco \ tends to decrease on
average. In \S~\ref{section:highredshift}, we detail specific
numbers for a sample merger that serves as an example for this
effect. 

We can see this more explicitly in
Figure~\ref{figure:xco_surfacedensity}, where we plot the mean \xco
\ versus mean surface density for the starburst snapshots in our
low-redshift sample of galaxies.  The starburst snapshots are defined
as the snapshots where the SFR peaks for a given model and are
categorised by the merger mass ratio (1:1, 1:3 and 1:10).  The mean
surface density is defined as the mass-weighted surface density over
all GMCs, $i$, in the galaxy:

\begin{equation}
  \langle \Sigma_{\rm H2}\rangle =\frac{\sum_{i} \Sigma_{\rm H2,i}
    \times M_{\rm H2,i}}{\sum_{i} M_{\rm H2,i}}
\label{eq:meansurfacedensity}
\end{equation}
Henceforth, when we refer to the surface density of the galaxy, we
refer to the mean surface density defined by
Equation~\ref{eq:meansurfacedensity}. We note that this is different
from the commonly used definition $\Sigma_{\rm H2} = M_{\rm H2}/A$,
where $A$ is the area within an observational aperture.  We refrain
from the latter definition as it is dependent on the choice of scale.
$\langle \Sigma_{\rm H2} \rangle$ can be thought of as the surface
density at which most of the mass resides.  Immediately, we see two
trends in Figure~\ref{figure:xco_surfacedensity}.

First, with increasing surface density, we see decreasing mean \xco
\ due to the warm and high velocity dispersion gas associated with
merging systems.  Second, the most violent mergers tend to have the
lowest mean \xco \ values, whereas lower mass ratio mergers (1:3) have
less extreme conditions, and thus $X$-factors more comparable to the
Galactic mean value.  The purple circles in
Figure~\ref{figure:xco_surfacedensity} note observational points from
\citet{tac08}; the models and data show broad agreement\footnote{We
  caution that our definition of $\langle \Sigma_{\rm H2} \rangle$ is
  a mass-weighted surface density, and is different from the surface
  density defined by the Tacconi et al. data, $M_{\rm H2}/A$.  In the
  limit of a large volume filling factor, these values will approach
  one another. The observed data closer to $\Sigma_{\rm mol} = 100
  \ \msun$pc$^{-2}$ likely represents galaxies with a clumpy ISM, and
  the modeled and observed \htwo \ gas surface densities may differ in
  this regime.}.

The open squares in Figure~\ref{figure:xco_surfacedensity} are points
from our models with mean metallicities less than $Z'<0.5$.  Careful
examination of these points shows that some of them have rather large
$X$-factors.  As we discussed in \S~\ref{section:metallicity}, lower
metallicity galaxies have larger mass fractions of CO-dark clouds, and
thus higher $X$-factors.

\begin{figure}
\hspace{-1cm}
\includegraphics[scale=0.4,angle=90]{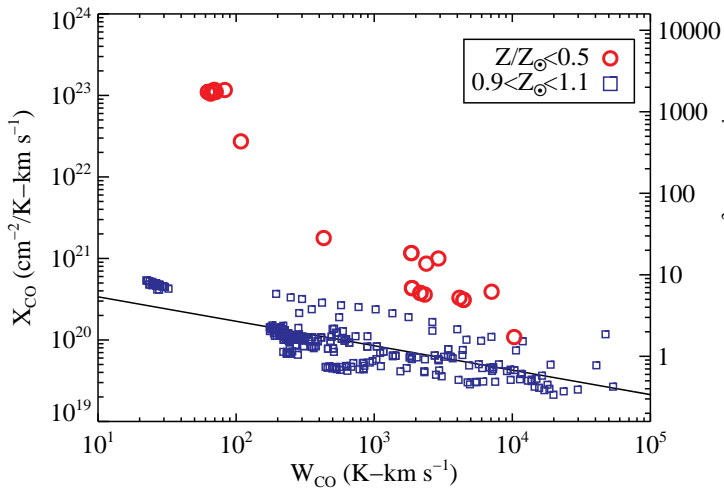}
\caption{\xco \ versus CO line intensity, $W_{\rm CO}$ for \z=0 1:1
  galaxy mergers in two distinct metallicity bins.  In this case,
  $W_{\rm CO}$ is a surrogate for $\Sigma_{\rm H2}$ that is an actual
  observable (and relateable via \xco).  At a given metallicity, \xco
  \ decreases with CO intensity due to the larger number of CO-dark
  GMCs in these galaxies.  The solid line shows the best fit model
  (c.f. \S~\ref{section:observations}) for a metallicity of
  $Z'=1$. \label{figure:xco_wco_mergers}}
\end{figure}

In Figure~\ref{figure:xco_wco_mergers}, we demonstrate more explicitly
the effect of metallicity on the \xco - surface density relationship.
Because we find it useful in a forthcoming section
(\S~\ref{section:observations}) to parameterise \xco \ in terms of the
observable velocity-integrated CO intensity, $W_{\rm CO}$ as an
observable, rather than $\Sigma_{\rm H2}$, we plot \xco \ against
$W_{\rm CO}$ in Figure~\ref{figure:xco_wco_mergers}.  $W_{\rm CO}$ is
defined as the luminosity-weighted line intensity from all GMCs in the
galaxy.  We plot the \xco-$W_{\rm CO}$ relationship for all snapshots
of all 1:1 \z=0 major mergers within two distinct metallicity
ranges. The selected metallicity ranges are arbitrary, and are chosen
simply to highlight the influence of metallicity.  As is clear, the
lowest metallicity points in Figure~\ref{figure:xco_wco_mergers} have
the highest $X$-factors, and the reverse is true for the highest
metallicity points.  For each metallicity bin, the trend is such that
increasing $W_{\rm CO}$ (or $\Sigma_{\rm H2}$) correlates with
decreasing \xco, though the normalisation varies with metallicity.
This informs our fitting formula in \S~\ref{section:observations}.

Returning to Figure~\ref{figure:xco_surfacedensity}, we make a final
point that there is a large dispersion in \xco \ for the merger
models.  Both the 1:3 models and 1:1 mergers show \xco \ values
ranging from above the Milky-Way mean to an order of magnitude below
it during their peak starburst.  When sampling the entire library of
merger orbits for a given merger mass ratio, a wide range in outcomes
is apparent.  We see a diverse set of velocity dispersions in the gas,
as well as star formation rates, owing to differing efficiencies at
which angular momentum is removed from the gas. Some models undergo
rather vigorous starbursts (approaching $\sim 500 $ \msunyrend),
whereas others hardly sustain a noticeable starburst upon final
coalescence.  Galaxies that undergo their peak starburst only on
first passage can have rather different metallicities in their ISM
than mergers that go through a vigorous star formation period during
first passage and inspiral before experiencing a starburst
contemporaneous with final coalescence. 

\begin{figure}
\hspace{-0.5 cm}
\includegraphics[angle=90,scale=0.4]{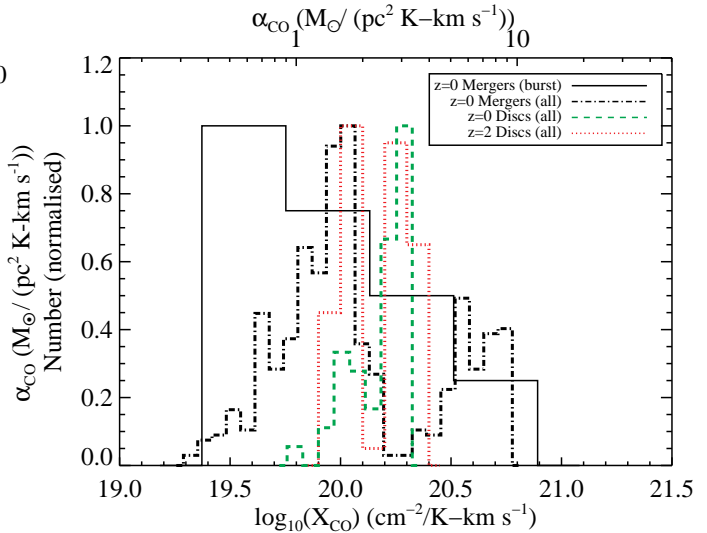}
\caption{Distribution of \xco \ values during the peak of the star
  formation rate (the ``burst'' snapshot) for all 1:1 and 1:3 local
  galaxy mergers (black solid), all \z=0 1:1 and 1:3 mergers (i.e. not
  just the burst snapshots; black dash-dot), all high-\z \ disc models
  with $Z'\approx 1$ (red dotted), and low-\z \ disc models with
  $Z'\approx 1 $ (green dashed).  See text for
  details.  \label{figure:xco_distribution}}
\end{figure}

 In the solid line of Figure~\ref{figure:xco_distribution}, we plot
 the distribution of emission-weighted mean \xco \ values for each of
 our 1:1 \z=0 merger models during the peak of their star formation
 rate. For comparison, we plot the distribution of $X$-factors for our
 \z=0 discs in the green dashed line (only plotting galaxies with
 $\sim$solar metallicity), as well as the distribution of \xco \ for
 our \z=2 discs (which we discuss in more detail in
 \S~\ref{section:highredshift}).  As we see, there is no ``merger''
 value for \xco.  It is possible to have \xco \ values in starbursting
 mergers comparable to the Milky Way's. The fact that starbursting
 mergers, on average, have lower $X$-factors than the Galactic mean is
 likely due to a selection effect. We return to this point in more
 (quantitative) detail in \S~\ref{section:observations}.

\subsection{\xco \ in High-Redshift Galaxies}
\label{section:highredshift}

\subsubsection{Basic Results}

Now that we have 
developed intuition regarding the variation of \xco \ with
metallicity and surface density, we are in a position to understand
how galaxies at high redshift may behave with respect to the CO-\htwo
\ conversion factor. 

Mergers at high-\z \ are some of the most luminous, rapidly
star-forming galaxies in the Universe.  As an example, many \zsim 2
Submillimetre-selected Galaxies (SMG) form stars at $\ga 1000$
\msun/yr \citep{nar10a}.  However, despite the $\sim$order of
magnitude greater star formation rate in these galaxies compared to
local mergers, mergers at high-\z \ are similar to their low-redshift
counterparts in terms of their typical $X$-factors.  While the mean
gas surface densities in e.g. \zsim 2 Submillimetre Galaxies (SMGs)
are larger than low-redshift mergers \citep[e.g.][]{tac08,nar10a},
both the dust temperatures and gas velocity dispersions also rise
commensurately.

As a specific example, we focus on a 1:1 major merger at \zsim 2 (This
is model ``z3b5e''; please refer to the Table in the Appendix for the
initial conditions of this model merger).  Model z3b5e undergoes a
luminous burst of $\sim 1500 $ \msunyrend, and may be selected as a
submillimetre galaxy when it merges \citep{nar09,hay11}.  During the
burst, this simulation reaches a mass-weighted mean surface density of
molecular gas of $\sim 10^4$ \msun pc$^{-2}$.  At the same time, the
mass weighted kinetic(dust) temperature is $\sim$150(160)
K\footnote{Note that owing to radiative transfer effects, this dust
  temperature is not necessarily what would be derived simply by
  identifying the location of the peak of the SED.}, and the
mass-weighted velocity dispersion in the GMCs is $\sim140$ \kmsend.
Doing a simple scaling results in a mean \xco \ of $\sim 5 \times
10^{19}$ \xcounits.  Of course the real mass-weighted value may vary
from this owing to both radiative transfer as well as the fact that
this simple scaling is not a true averaging.  Observational estimates
of the $X$-factor in SMGs suggest that they are similar to the lower
values in the range of local ULIRGs \citep{tac08}.

Our simulated disc galaxies at high-redshift show a range of \xco
\ values, ranging from comparable to the Galactic mean to values 2-5
times lower.  The reason massive discs at high-redshift may have lower
$X$-factors than the Galactic mean can be understood in the following
way.  In contrast to present-epoch galaxies, galaxies at
higher-redshifts (\z$\ga 1$) at a fixed stellar mass are denser and
more gas-rich \citep[e.g .][]{erb06a,for09,dad10a,tac10}.  Both
simulations and observations suggest that galaxies around redshifts
\zsim1-2 may have baryonic gas fractions of order 20-60\%
\citep{dav10,dad10a,tac10}.  A primary consequence of this is that
discs at higher-redshifts may be heavily star-forming, with star
formation rates or order $\sim10^2$ \msunyrend, comparable to local
galaxy mergers \citep{dad07,for09,nar10b}.  In fact, simulations
suggest that disc galaxies at \zsim 2 likely dominate the infrared
luminosity function \citep{hop10}.

In the absence of rather extreme stellar feedback, very gas-rich discs
at high-redshift can be unstable to fragmentation, and form massive
$\sim$kpc-scale clumps \citep[e.g.][]{spr05a,cev10}.  These clumps
can have relatively high velocity dispersions ($\sim10^2$
\kmsend) and warm gas temperatures owing to high volumetric densities
and high star formation rates \citep{bou10}.

These effects are the strongest in the most massive discs.  Our most
massive \z=2 model disc galaxy has a total baryonic mass of $\sim
5\times 10^{11}$ \msunyrend, and has typical\footnote{Because our
  simulations are not cosmological, there is no accretion of
  intergalactic gas.  As a result, the metallicities in our model
  galaxies only rise with time.  Because the $X$-factor is dependent
  on metallicity (\S~\ref{section:metallicity}), we have to make a
  choice as to which snapshot/metallicity to consider as a 'typical'
  galaxy.  We assume any snapshot above $Z'>0.5$ is ``typical'' based
  on the steady-state metallicities found for galaxies of baryonic
  mass comparable to those in our sample from cosmological modeling
  \citep[Figure 2 of ][though see \citet{ker11,vog11} and
    \citet{sij11}]{dav10}.} $X$-factors ranging anywhere between a
factor of five below the Galactic mean to the Galactic mean value.
The lower mass \z=2 isolated disc models (with baryonic masses of
$M_{\rm bar} = 1\times10^{11}$ and $3.5\times10^{10}$ \msunend)
typically have $X$-factors comparable to the Milky Way mean.
Returning to Figure~\ref{figure:xco_distribution}, we examine the red
dotted line that represents \zsim 2 disc models. Because the idea of a
'starburst' snapshot is less meaningful for the evolution of a disc
galaxy, we plot the $X$-factor for every snapshot for our model discs
with metallicities around solar.  We see a large spread in mean
$X$-factors.

\subsubsection{Do Mergers and Discs Have Inherently Different $X$-factors?}

In light of the fact that high-\z \ discs have, at times, star
formation rates comparable to local galaxy mergers, a pertinent
question is whether there is an intrinsic difference in the $X$-factor
between high-\z \ discs and galaxy mergers. Another way of saying
this is, for a given set of physical conditions, are the $X$-factors
from mergers lower than the $X$-factors from high-\z \ gas-rich,
gravitationally unstable discs?  A cursory examination of
Figure~\ref{figure:xco_distribution} indicates that mergers (the
black solid line) have systematically lower $X$-factors than discs
(the blue and red dashed and dotted lines).  Indeed, in the local
Universe, it is observed that mergers have, on average, lower
$X$-factors than discs \citep[e.g.][]{tac08}.  However, this is
likely due to a selection bias.  We remind the reader that the black
solid line in Figure~\ref{figure:xco_distribution} represents {\it
  starbursting} mergers.  These mergers are caught when their gas is
extremely warm and with large velocity dispersion.  When comparing
mergers and discs with comparable physical conditions, the observed
\xco \ values are in fact quite similar.  It is the physical conditions
in a galaxy that determine the $X$-factor, not the global morphology.

 To demonstrate this, we perform three tests.  First, we compute the
 distribution of \xco \ values for {\it all} 1:1 and 1:3 merger
 snapshots (at $\sim$solar metallicity), and indicate this with the
 black dot-dashed line in Figure~\ref{figure:xco_distribution}.  As we
 see, the distribution of \xco \ values is broad, but with
 substantially less power in the low $X$-factor regime than the
 distribution that denotes only starbursting mergers (black solid
 line).  This highlights that mergers which are selected during a
 particularly active phase are more likely to have low $X$-factors,
 due to their warm and high-$\sigma$ gas.  When controlling for this
 effect by picking galaxies with similar CO intensity ($W_{\rm CO}$)
 and metallicity, mergers and discs have the same \xco \ on average.

\begin{figure}
\includegraphics[scale=0.4,angle=90]{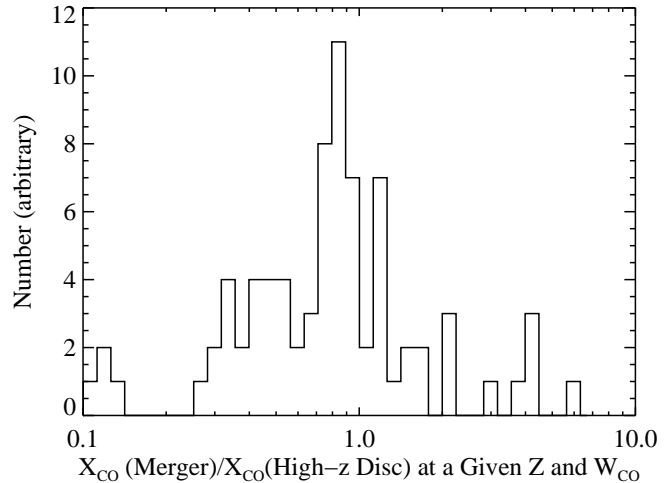}
\caption{Comparison of the $X$-factor between low-\z \ mergers (1:1
  and 1:3) and high-\z \ star forming discs.  The histogram denotes
  ratio of $X$-factor from mergers versus high-\z \ discs between
  snapshots with a similar metallicity and CO intensity.  The sharp
  peak near unity implies that galaxies with similar physical
  conditions have similar $X$-factors, independent of large-scale morphology.
  \label{figure:mergers_highzdiscs}}
\end{figure}

To show this, in Figure~\ref{figure:mergers_highzdiscs}, we perform
our second test in which we examine the $X$-factors from all the 1:1
and 1:3 mergers (at low \z) and compare them to the \xco \ from
high-\z \ discs with the same\footnote{``The same'' here means that
  the values of $Z'$ and $W_{\rm CO}$ are within 10\% of one another.}
metallicity and CO intensity ($W_{\rm CO}$).  We could equivalently
perform this analysis in terms of $\Sigma_{\rm H2}$, though as we will
show in \S~\ref{section:observations}, parameterising in terms of
$W_{\rm CO}$ is desirable with regards to observations.  There is a
strong peak at \xco \ ratios near unity, with some spread.  The median
value in the distribution is $\sim 0.8$, and the mean is $\sim 1.1$.
The implication from Figure~\ref{figure:mergers_highzdiscs} is that
galaxies with similar physical conditions (here, $Z'$ and $W_{\rm
  CO}$) have similar $X$-factors, regardless of whether they are discs
or mergers.  The fact that mergers, on average, have lower $X$-factors
than discs in the local Universe likely derives from the fact that
they are selected as starbursts, which have preferentially higher
temperatures and velocity dispersions in the gas.

Third, in Figure~\ref{figure:xco_surfacedensity_highz}, we examine the
relationship between \xco \ and $W_{\rm CO}$ for the same galaxies
plotted\footnote{To reduce clutter in the plot, we randomly draw 10\%
  of the galaxies within each merger ratio bin to plot.} in
Figures~\ref{figure:xco_distribution} and
\ref{figure:mergers_highzdiscs}. These are all galaxies with
metallicities around solar.  The principal result from
Figure~\ref{figure:xco_surfacedensity_highz} is that galaxies within a
relatively limited metallicity and $W_{\rm CO}$ (or surface-density)
range have similar $X$-factors, regardless of the type of merger it
is. Mergers and discs have similar \xco \ values when they have
similar physical conditions, and are not inherently different based on
their global morphology.  In addition,
Figure~\ref{figure:xco_surfacedensity_highz}, like
Figures~\ref{figure:xco_surfacedensity} and
\ref{figure:xco_wco_mergers}, shows a systematic decrease of \xco
\ with increasing $W_{\rm CO}$ (and $\Sigma_{\rm mol}$).

\section{Application to Observations}
\label{section:observations}
\subsection{Deriving \xco \ from Observations}

As we have seen from the previous sections, it is clear that there is
a continuum of \xco \ values that vary with galactic environment.
The dominant drivers of the $X$-factor in our simulations are the
metallicity of the star-forming gas, and the thermal and dynamical
state of the GMCs. 
Informed by this, we are motivated to parameterise \xco \ as a
function of observable properties of galaxies.

Metallicity is a crucial ingredient to any parameterisation.  At
subsolar metallicities, we see the rapid growth of CO-dark GMCs.  This
has been noted both in observations \citep[e.g.][]{ler11,gen11b}, as
well as other numerical models \citep[][]{she11b,kru11a,fel11b}.  As we saw in
\S~\ref{section:metallicity}, as well as in
Figure~\ref{figure:xco_wco_mergers}, at a given galaxy surface density
(or CO intensity), \xco \ increases with decreasing metallicity.

Beyond this, as was shown in \S~\ref{section:surfacedensity}, as well
as in the models of \citet{nar11b}, galaxy surface density is
correlated with the thermal and dynamical state of the gas: at a given
metallicity, higher surface density galaxies, on average, correspond
to galaxies with a warm and high velocity dispersion molecular ISM, due
to their higher SFRs.

\begin{figure}
\hspace{-1cm}
\includegraphics[scale=0.4,angle=90]{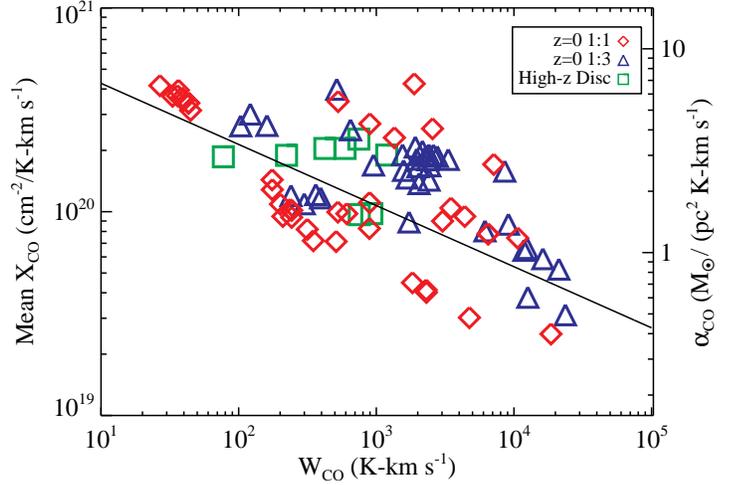}
\caption{Comparison of \xco \ versus CO intensity ($W_{\rm CO}$) for
  low-\z \ galaxy mergers and high-\z \ discs in an effort to
  investigate if mergers and discs inherently have different \xco
  \ properties.  Included in this plot are all 1:1 and 1:3 mergers
  simulated at \z=0.  Only snapshots with metallicities $Z'>0.7$ are
  shown. To reduce clutter in the plot, we plot only a randomly drawn
  subsample (10\%) of the snapshots from each mass ratio.  The line
  shows the best fit from Equation~\ref{eq:xcozsd_ico_lazy}.
  Evidently, galaxies that have similar physical conditions have
  similar $X$-factors, independent of galaxy morphology or
  evolutionary history. See text for
  details. \label{figure:xco_surfacedensity_highz}}
\end{figure}

Informed by these results, we perform a 2D Levenberg-Marquardt fit
\citep{mar09b} on our model galaxies (considering every snapshot of
every model), fitting \xco \ as a function of mass-weighted mean
metallicity and mass-weighted mean \htwo \ surface density.  We find
that our simulation results are reasonably well fitted by a function of
the form:
\begin{equation}
\label{eq:xcozsd}
\xco \approx \frac{1.3\times10^{21}}{Z' \times \langle\Sigma_{\rm
    H2}\rangle^{0.5}}
\end{equation}
where $\Sigma_{\rm H2}$ is in units of \msun pc$^{-2}$ and \xco \ is
in units of \xcounits.  Equation~\ref{eq:xcozsd} provides a good fit
to the model results above metallicities of $Z' \approx 0.2$. Turning
again to
Figures~\ref{figure:xco_metallicity},~\ref{figure:xco_surfacedensity},~\ref{figure:xco_wco_mergers}
and ~\ref{figure:xco_surfacedensity_highz}, we highlight the solid
lines which show how Equation~\ref{eq:xcozsd} fits both the simulation
results and observational data.  We note that \citet{ost11} obtained a
similar result, $\alpha_{\rm CO} \propto \Sigma_{\rm mol}^{-0.5}$ by
interpolating between empirical $\alpha_{\rm CO}$ values ($\alpha_{\rm
  CO}=3.2$ for $\Sigma_{\rm mol}=100$ \msun pc$^{-2}$ and $\alpha_{\rm
  CO}=1$ for $\Sigma_{\rm mol} = 1000$ \msun pc$^{-2}$).

Because $\Sigma_{\rm H2}$ is not directly observable (hence the need
for an $X$-factor), we re-cast Equation~\ref{eq:xcozsd} in terms of
the velocity-integrated CO line intensity.  In order to
parameterise \xco \ in a manner that is independent of the effects
of varying beam-sizes or observational sensitivity, we define the
observable CO line intensity as the luminosity-weighted CO intensity
over all GMCs, $i$:
\begin{equation}
\langle W_{\rm CO}\rangle = \frac{\int W_{\rm CO}^2 \ dA}{\int W_{\rm
    CO} \ dA} \equiv \frac{\sum{L_{\rm CO,i} \times W_{\rm
      CO,i}}}{\sum{L_{\rm CO,i}}}
\end{equation}
where $\langle W_{\rm CO} \rangle$ is in units of K-\kmsend, and is
the CO surface brightness of the galaxy.  We then fit to obtain a
relation between \xco, $Z'$, and $\langle W_{\rm CO} \rangle$:

\begin{equation}
\label{eq:xcozsd_ico_lazy}
\xco = \frac{6.75\times10^{20} \times \langle W_{\rm CO}\rangle^{-0.32}}{Z'^{0.65}}
\end{equation}
where again $\langle W_{\rm CO} \rangle$ is CO line intensity measured
in K-\kms, \xco \ is in \xcounits, and $Z'$ is the metallicity divided
by the solar metallicity.  By converting \xco \ to $\alpha_{\rm CO}$, we similarly obtain:

\begin{equation}
\alpha_{\rm CO} = \frac{10.7 \times \langle W_{\rm CO} \rangle^{-0.32}}{Z'^{0.65}}
\end{equation}
Where $\alpha_{\rm CO}$ is in units of \msun ${\rm
  pc}^{-2}$(K-\kmsend)$^{-1}$.

It is important to recognize that the power-law in
Equation~\ref{eq:xcozsd_ico_lazy} cannot describe \xco \ indefinitely.
At very low $W_{\rm CO}$, GMCs tend toward fixed properties in
galaxies and the galactic environment plays a limited role in setting
\xco.  Considering this, Equation~\ref{eq:xcozsd_ico_lazy} formally
becomes:

\begin{equation}
\label{eq:xcozsd_ico_formal}
\xco = \frac{{\rm min}\left[4,6.75 \times \langle W_{\rm CO}
    \rangle^{-0.32}\right] \times 10^{20}}{Z'^{0.65}}
\end{equation}
or, similarly:
\begin{equation}
\label{eq:xcozsd_ico_formal2}
\alpha_{\rm CO} = \frac{{\rm min}\left[6.3,10.7 \times \langle W_{\rm CO}\rangle^{-0.32} \right]}{Z'^{0.65}}
\end{equation}

Equations~\ref{eq:xcozsd_ico_formal} and \ref{eq:xcozsd_ico_formal2}
can be used directly with observations of galaxies to infer an
expected $X$-factor.  One advantage of this formalism is that it
captures the continuum of CO-\htwo \ conversion factors, rather than
utilising bimodal ``Disc'' and ``ULIRG'' values.  Because we have
chosen the physical quantities in our modeling based on mass or
luminosity-weighted averages, they are defined without reference to a
particular scale.  Consequently, Equation~\ref{eq:xcozsd_ico_formal}
can be used on scales ranging from our resolution limit of $\sim 70$
pc to unresolved observations of galaxies.

It is conceivable that alternative definitions of the observed mean CO
intensity could be appropriate.  One can imagine implementing an
area-weighted intensity, i.e. $W_{\rm CO} = L_{\rm CO}$/Area.  This
has the undesirable attribute of being dependent on a defined scale.

\subsection{The Kennicutt-Schmidt Star Formation Relation in Galaxies from \z=0-2}
\label{section:kslaw}

A natural application of our model for the CO-\htwo \ conversion
factor is the Kennicutt-Schmidt star formation rate surface density-
gas surface density relation in galaxies.  Because the inferred \htwo
\ gas masses from observed galaxies are inherently dependent on
conversions from CO line intensities, our understanding of the
Kennicutt-Schmidt relation is fundamentally tied to the potential
variation of \xco \ with the physical environment in galaxies.

Recent surveys of both local galaxies \citep[e.g.][and references
  therein]{ken98a,ken98b,big08}, as well as pioneering efforts at
higher redshifts \citep[e.g.][]{bou07,bot09,dad10b,gen10} have
provided a wealth of data contributing to our knowledge of the star
formation relation in both quiescent disc galaxies and starbursts.
Work by \citet{dad10b} and \citet{gen10} demonstrate the sensitivity
of these relations to the CO-\htwo \ conversion factor: When applying
the traditional bimodal conversion factor (\xco $ \approx 8\times
10^{19} $ \xcounits \ for ULIRGs and \xco $ \approx 2\times10^{20} $
\xcounits \ for discs) to the starburst galaxies and discs,
respectively, a bimodal SFR relation becomes apparent when the data
are plotted in terms of $\Sigma_{\rm SFR}$ and $\Sigma_{\rm mol}$.

\begin{figure*}
\includegraphics[angle=90,scale=0.8]{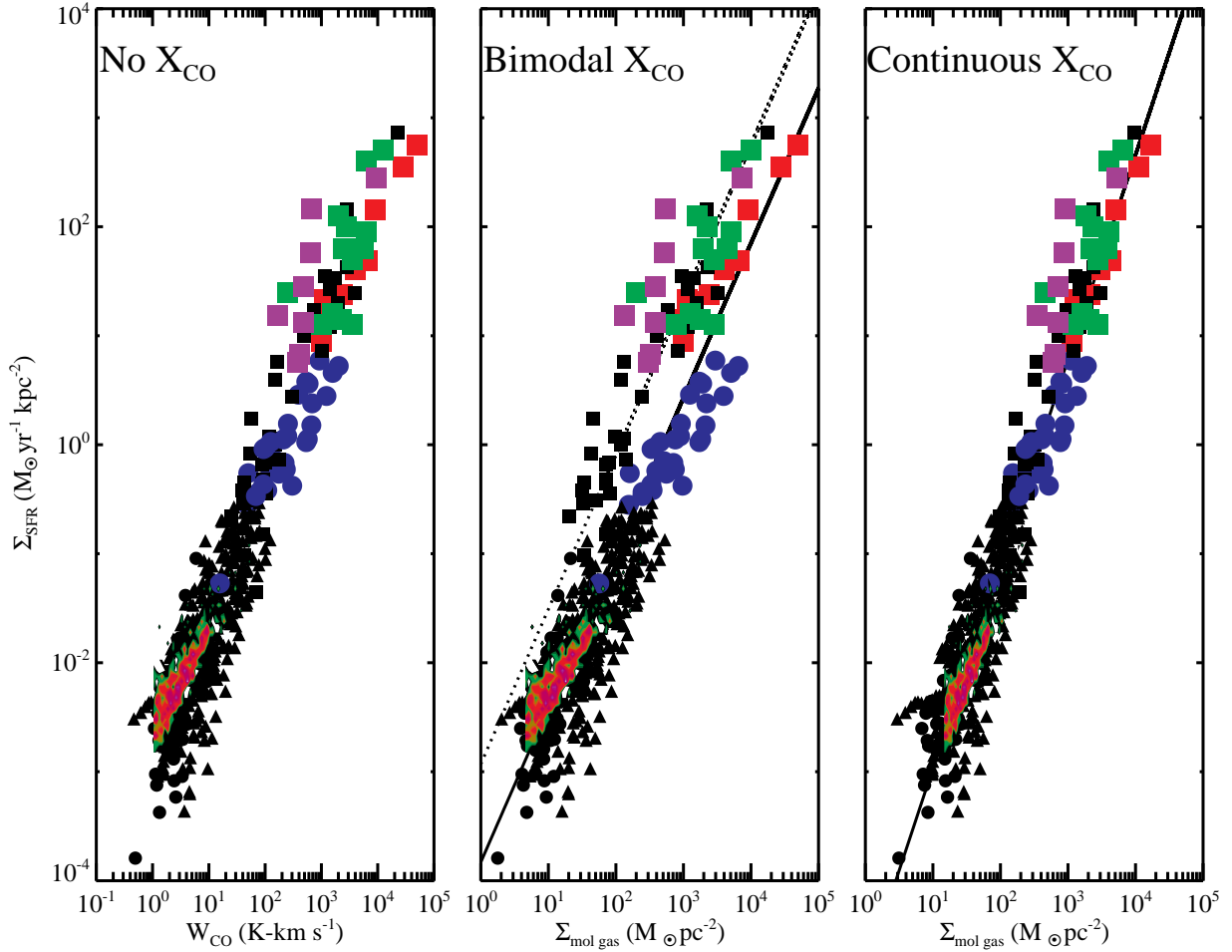}
\caption{Kennicutt-Schmidt star formation relation (SFR surface
  density versus \htwo \ gas surface density) in observed galaxies.
  Circles and triangles are local discs or high-\z \ \bzk \ galaxies,
  and squares are inferred mergers (local ULIRGs or high-\z \ SMGs).
  Colours denoting separate surveys are described below.  {\it Left}:
  SFR surface density vs. velocity-integrated CO intensity, yielding a
  unimodal SFR relation.  {\it Centre}: When applying an effectively
  bimodal \xco \ ($\alpha_{\rm CO}= 4.5$ for local discs, 3.6 for
  high-\z \ discs, and 0.8 for mergers), the resulting SFR relation is
  bimodal.  The solid and dotted lines overplotted are the best fit
  tracks for each ``mode'' of star formation as in
  \citet{dad10b}. {\it Right}: SFR relation when applying
  Equation~\ref{eq:xcozsd_ico_formal} to the observational data,
  resulting in a unimodal SFR relation.  The power-law
  index in the relation is approximately 2 (solid line).  {\it Symbol
    legend}: We divide galaxies into 'disc-like' with
  filled circles, and 'merger-like' with squares.  This assumes that
  high-\z \ \bzk \ galaxies are all discs, high-\z \ SMGs and low-\z
  \ ULIRGs are all mergers. The low-\z \ disc observations (black
  filled circles and black triangles) come from
  \citet{ken07,won02,cro07,sch07} and comprise both resolved and
  unresolved points. The resolved points from the survey of
  \citet{big08} are denoted by the coloured contours.  The local
  ULIRGs are compiled by \citet{ken98b} and are denoted by black
  filled squares.  The high-\z \ discs come from \citet{gen10},
  \citet{dad10a,dad10b} and are represented by filled blue circles.
  The high-\z \ SMGs are divided into the the samples of \citet{bot09}
  (purple), \citet{bou07} (green), and \citet{gre05,tac06,tac08,eng10}
  as compiled by \citet{gen10} (filled red
  squares). \label{figure:ks}}
\end{figure*}

\begin{figure*}
\includegraphics[angle=90,scale=0.8]{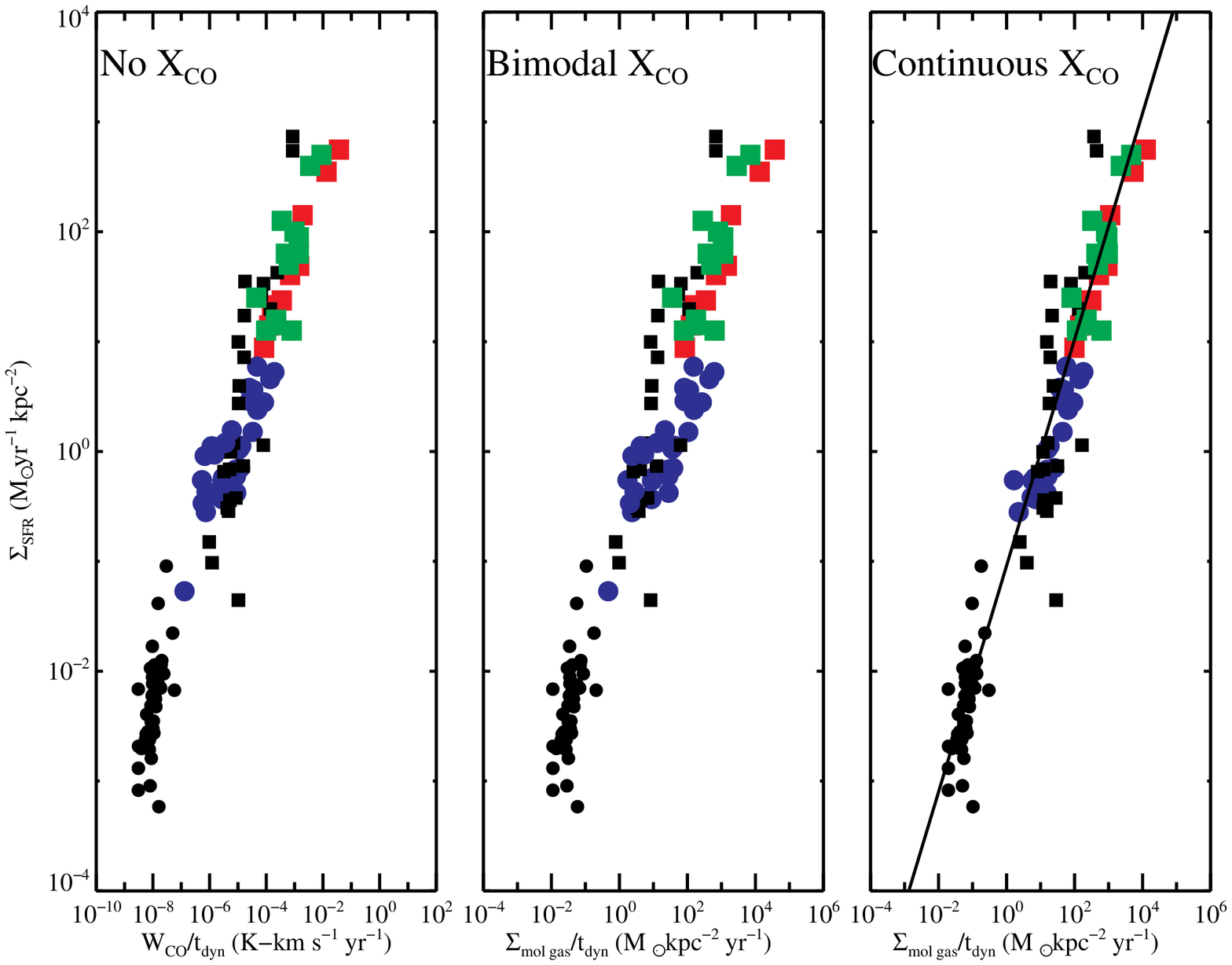}
\caption{Similar to Figure~\ref{figure:ks}, but with the abscissa
  showing $W_{\rm CO}$ or $\Sigma_{\rm mol}$ divided by the orbital
  time of the observed galaxy.  Symbols are the same as in
  Figure~\ref{figure:ks}, but we omit galaxies for which orbital times
  are not available.  The best fit slope in the right panel is of
  order unity.  See text for details. \label{figure:ks_tdyn}}
\end{figure*}

\begin{figure*}
\includegraphics[angle=90,scale=0.8]{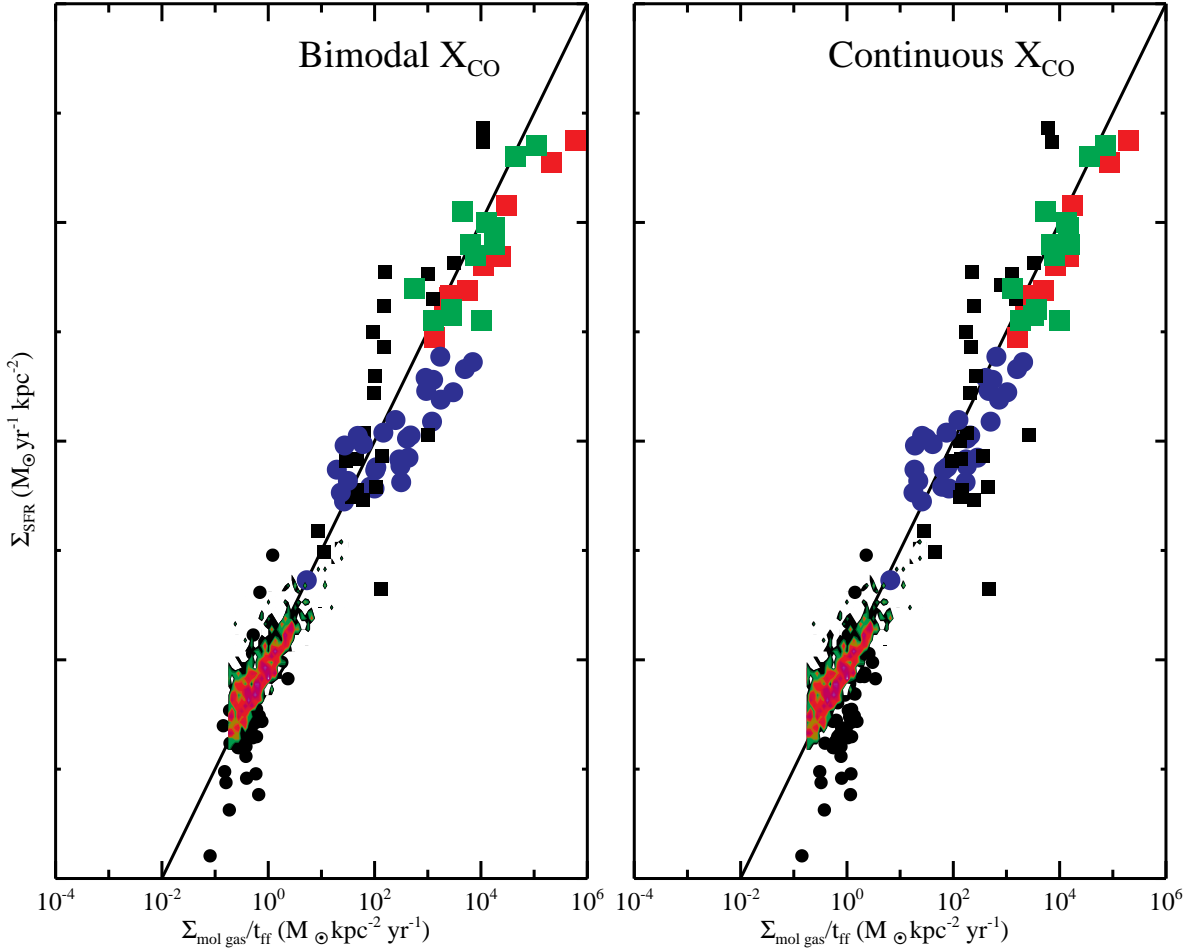}
\caption{ Same as Figure~\ref{figure:ks}, but instead of plotting
  $\Sigma/t_{\rm dyn}$ on the abscissa we instead plot $\Sigma/t_{\rm
    ff}$, where $t_{\rm ff}$ is the free-fall time in the star-forming
  clouds of a galaxy -- see main text for details. We do not include a
  $\Sigma_{\rm SFR}-W_{\rm CO}/t_{\rm ff}$ plot as the calculation of
  $t_{\rm ff}$ requires a gas mass. Symbols are the same as in
  Figures~\ref{figure:ks} and \ref{figure:ks_tdyn}. The left panel
  shows the results using a bimodal \xco, while the right panel shows
  the results using our continuous \xco. The solid black lines show
  the relation $\Sigma_{\rm SFR} = \epsilon_{\rm ff} \Sigma_{\rm
    mol}/t_{\rm ff}$ with $\epsilon_{\rm ff} = 0.01$.
\label{figure:tff_ks}}
\end{figure*}

In Figures \ref{figure:ks} -- \ref{figure:tff_ks}, we illustrate the
effects of our model fit for \xco \ as a function of galaxy physical
properties.  In the left panel of Figure~\ref{figure:ks}, we plot the
star formation rate surface density for both local galaxies and
high-\z \ galaxies as compiled by \citet{dad10b} and \citet{gen10}
against their CO line intensity, $W_{\rm CO}$.  Although there is
significant scatter, the $\Sigma_{\rm SFR}$ vs $W_{\rm CO}$ relation
is unimodal.  In the middle panel, we plot the SFR-$\Sigma_{\rm H2}$
relation utilising the bimodal $X$-factors assumed in the literature
(with the above ``ULIRG'' value for the inferred mergers [local ULIRGs
  and high-\z \ SMGs], and the above ``quiescent'' value for low-\z
\ discs and high-\z \ \bzk \ galaxies)\footnote{In practice, for the
  high-\z \ \bzk \ galaxies, we utilise an $X$-factor of \xco $\approx
  2.3 \times 10^{20}$ \xcounits \ to remain consistent with
  \citet{dad10b}, though the usage of this versus the more standard
  disc value makes little difference.}.  The circles are unresolved
observations of disc galaxies at low and high-\z, triangles and
contours are resolved observations of local discs, and the squares are
local ULIRGs and inferred mergers at high-redshift.  When separate
high and low values of \xco \ are adopted for discs and mergers, a
bimodal relationship between $\Sigma_{\rm SFR}$ and $\Sigma_{\rm gas}$
results, with power-law index ranging between unity and 1.5.

In the right panel of Figure~\ref{figure:ks}, we apply
Equation~\ref{eq:xcozsd_ico_formal} to the observational data
(assuming $Z'=1$ for the galaxies).  The scatter in the modified
relation immediately tightens, and it becomes unimodal.  To
numerically quantify the reduction in scatter with the modified
relation we examine the ratio of the maximum inferred $\Sigma_{\rm
  H2}$ to the minimum for all points within the relatively tight SFR
surface density range of [0.05,0.1] \msunyr kpc$^{-2}$ for the centre
and right panels of Figure~\ref{figure:ks}.  The scatter is reduced by
approximately a factor of 5.  Within this $\Sigma_{\rm SFR}$ range, no
mergers are in the sample.  Thus, the reduction in scatter is not due
to simply using a unimodal \xco \ versus bimodal \xco.  We note that
in applying Equation~\ref{eq:xcozsd_ico_formal} to the observed data,
we have to assume that the intensity within the reported area is
uniform.  If the emission is instead highly concentrated over very few
pixels, then the application of Equation~\ref{eq:xcozsd_ico_formal}
may overestimate \xco.

The reason for the transition from a bimodal to unimodal KS relation
is clear.  In the modified relation, similar to the traditional
Kennicutt-Schmidt plot that uses a bimodal $X$-factor, the lower
luminosity discs have CO-\htwo \ conversion factors comparable to the
Galactic mean, and the most luminous discs have $X$-factors up to an
order of magnitude lower.  Very massive, gas-rich, unstable discs, as
well as lower luminosity mergers have $X$-factors in between the two,
however, and fill in the continuum. Utilising
Equation~\ref{eq:xcozsd_ico_formal}, a simple linear chi-square fit of
the observed data on the right side of Figure~\ref{figure:ks} returns:
\begin{equation}
\label{eq:ks}
{\rm log_{\rm 10}}(\Sigma_{\rm SFR}) = 1.95 \times {\rm log_{\rm 10}}(\Sigma_{\rm
  mol}) - 4.9
\end{equation}
where $\Sigma_{\rm SFR}$ is measured in $\msunyr {\rm kpc}^{-2}$ and
$\Sigma_{\rm mol}$ is measured in \msun pc$^{-2}$. Utilising an
empirical method to obtain \xco $\propto W_{\rm CO}^{-0.3}$,
\citet{ost11} previously showed that the observational data compiled
in \citet{gen10} yields a similar fit to Equation~\ref{eq:ks}.

We remind the reader of the assumptions that have gone into this fit:
We have assumed that every galaxy has solar metallicity, and
neglected any potential effects of differential excitation in the CO
as a function of infrared luminosity
\citep[e.g.][]{nar11a}. Nevertheless, the application of a variable
$X$-factor on the SFR-gas surface density relation has interesting
implications. 

 First, the index of $\sim 2$ of Equation~\ref{eq:ks} is consistent
 with the analytic models and hydrodynamic simulations of
 self-regulated star formation by \citet{ost11}.  This work suggests
 that in molecular regions where supernova-driven turbulence controls
 the SFR and gas dominates the vertical gravity, the SFR surface
 density should be proportional to the gas surface density squared: $
 {\rm log} (\Sigma_{\rm SFR}) = 2 \times {\rm log} (\Sigma_{\rm mol})
 - 5.0$ (adopting fiducial parameters in Equation 13 of
 \citet{ost11}).  This is shown as the solid line in the right panel
 of Figure~\ref{figure:ks}, and is very comparable to the best fit
 relation.  Second, comparing Equation~\ref{eq:ks} to Equation 13 of
 \citet{ost11}, the empirical results are consistent with a value of
 momentum injected/total stellar mass formed of $f_p \times p_*/m_*
 \sim 3000 \ \kms (\Sigma_{\rm mol}/100 \ \msun {\rm
   pc}^{-2})^{-0.05}$; the fiducial value adopted in \citet{ost11} is
 3000 \kmsend.

\begin{figure*}
\includegraphics[angle=90,scale=0.35]{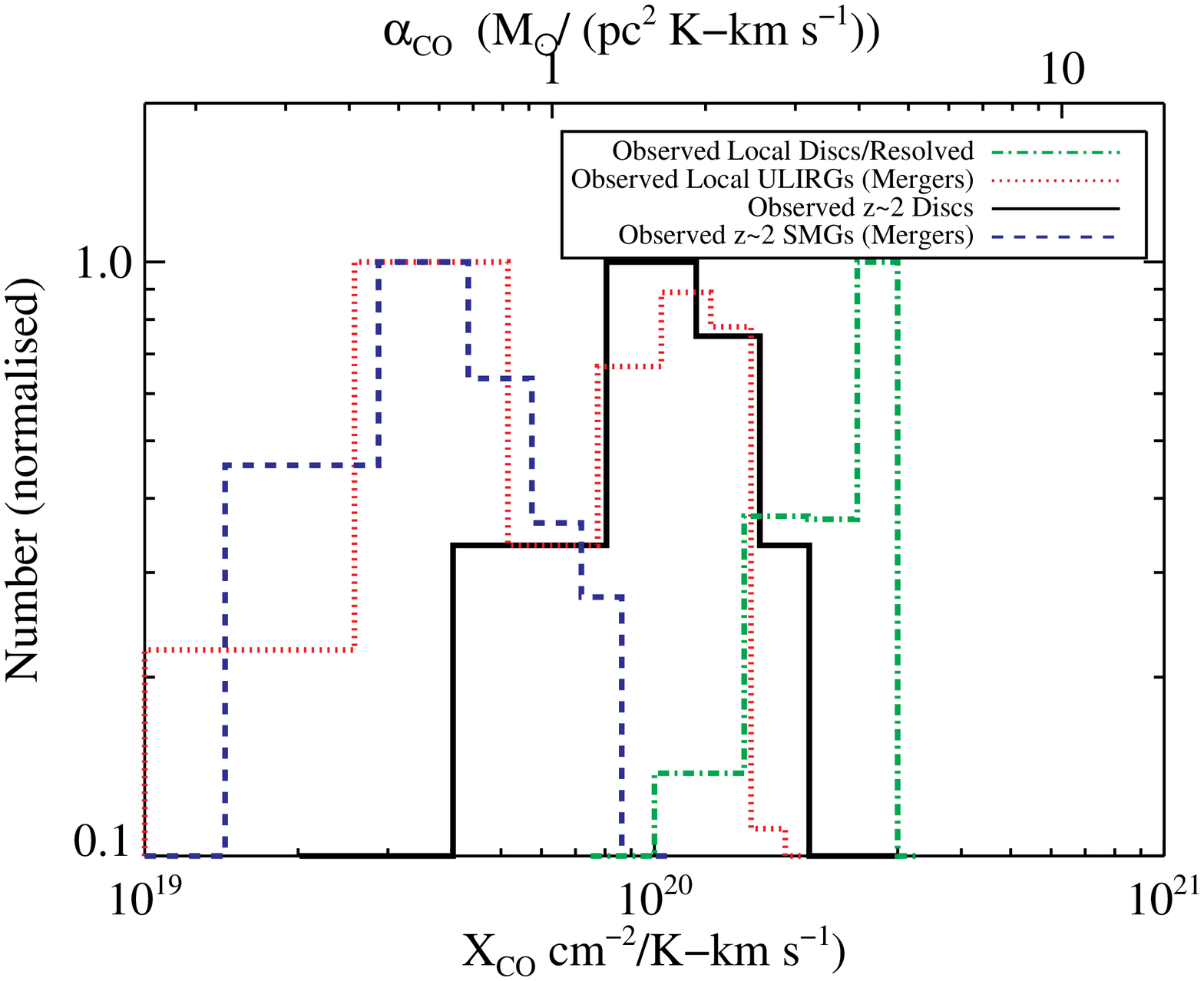}
\includegraphics[angle=90,scale=0.35]{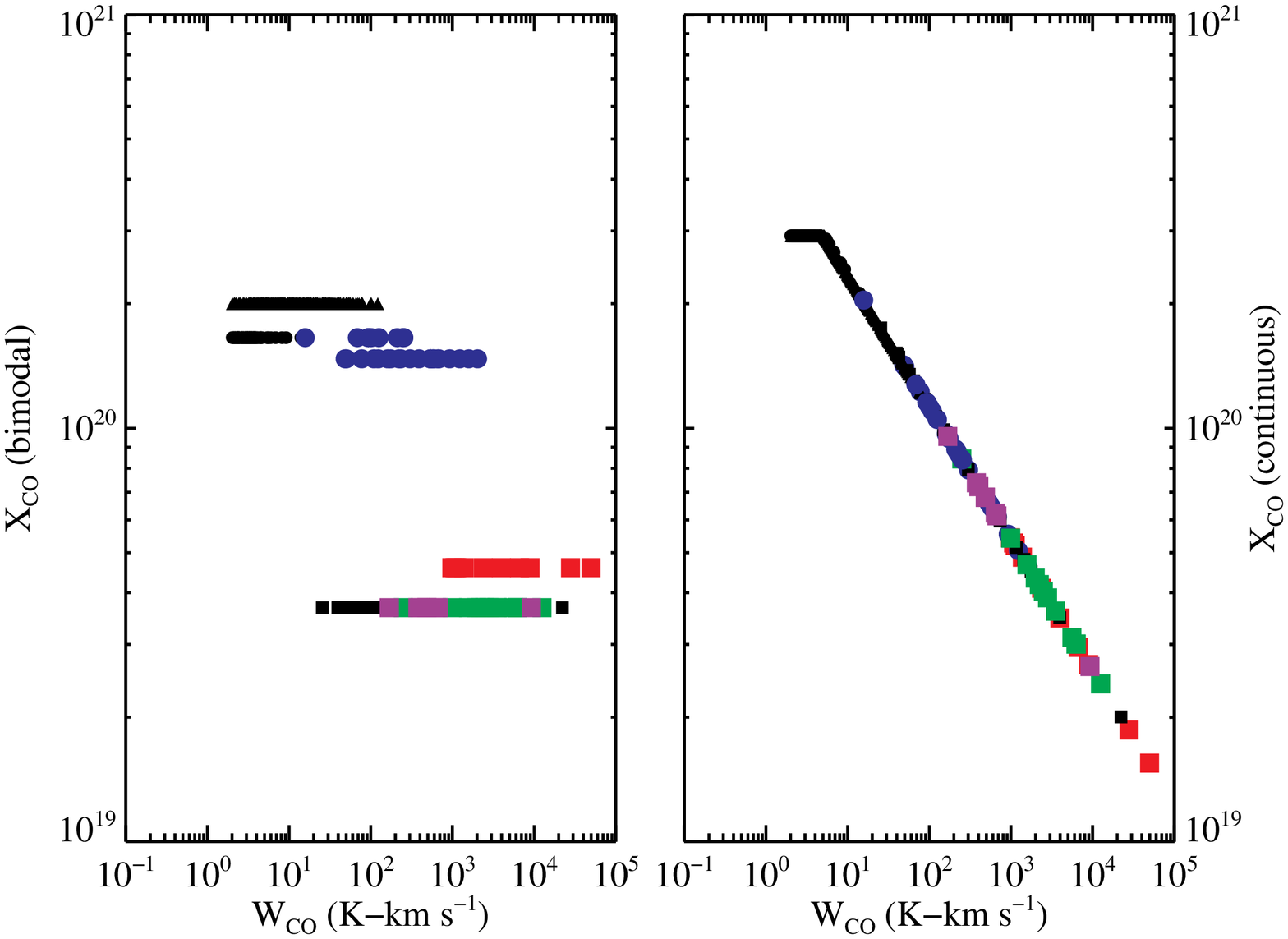}
\caption{{\it Left:} We utilise our model fit for \xco
  (Equation~\ref{eq:xcozsd_ico_formal}) to infer the \xco \ values for
  observed galaxies compiled by \citet{gen10} and \citet{dad10b}.
  {\it Right:} CO line intensity, $W_{\rm CO}$ versus \xco \ for the
  observed galaxies from Figure~\ref{figure:ks} (the colour coding of
  the points is the same as in Figure~\ref{figure:ks}).  The \xco
  \ values are both the original literature values (denoted by
  ``\xco \ (bimodal)'', and our derived values (denoted by ``\xco
  \ (continuous)'').  \label{figure:bzkulirg}}
\end{figure*}

\citet{dad10b} and \citet{gen10} suggest an alternative mechanism for
reducing the scatter imposed by the utilisation of a bimodal \xco \ in
Figure~\ref{figure:ks}.  Specifically, these authors find that by
dividing the molecular gas surface density by the galaxy's orbital
time, the observed Kennicutt-Schmidt relation goes from bimodal to
unimodal, suggesting that the galaxy's global properties are related
to the local, small-scale processes of star formation.  That is to
say, when using a bimodal \xco, the $\Sigma_{\rm SFR}-\Sigma_{\rm
  mol}$ relationship is bimodal, whereas the $\Sigma_{\rm
  SFR}-\Sigma_{\rm mol}/t_{\rm dyn}$ relationship is unimodal, with
some scatter.  

If one abandons the bimodal \xco \ approximation, and utilises our
favoured model for \xco, the observed relationship between
$\Sigma_{\rm SFR}$ and $\Sigma_{\rm mol}/t_{\rm dyn}$ remains
unimodal, and in fact the scatter in the relation is reduced compared
to what one obtains using a bimodal \xco.  To show this, in
Figure~\ref{figure:ks_tdyn}, we show the analog to
Figure~\ref{figure:ks}, but with the abscissa showing the surface
density or $W_{\rm CO}$ divided by the dynamical time.  The dynamical
times used are the same as those in \citet{dad10b} and \citet{gen10},
and are defined as the rotational time at either the galaxy's outer
radius, or half-light radius, depending on the sample.  In the left
panel of Figure~\ref{figure:ks_tdyn}, we show the relationship between
$\Sigma_{\rm SFR}$ and $W_{\rm CO}/t_{\rm dyn}$ (i.e. pure
observables); in the centre panel, we show the relationship between
$\Sigma_{\rm SFR}$ and $\Sigma_{\rm mol}/t_{\rm dyn}$ when assuming a
bimodal \xco \ (as is done in \citet{dad10b} and \citet{gen10}), and
in the right panel we show the same relationship, but with
$\Sigma_{\rm mol}$ determined using our best-fit continuous \xco,
rather than the bimodal \xco \ value used in \citet{dad10b} and
\citet{gen10}.  We find a best fit relation (using our model for \xco)
of:
\begin{equation}
{\rm log_{10}} (\Sigma_{\rm SFR})= 1.03 \times {\rm
  log_{10}}(\Sigma_{\rm mol}/t_{\rm dyn})-1.05
\end{equation}
where $\Sigma_{\rm SFR}$ and $\Sigma_{\rm mol}/t_{\rm dyn}$ are both
in $\msunyr {\rm kpc}^{-2}$.

 Our best fit $\Sigma_{\rm SFR}-\Sigma_{\rm mol}/t_{\rm dyn}$ relation
 has slope of approximately unity, comparable to what is found by
 \citet{gen10}, and consistent with the best fit slope of
 \citet{dad10b} of $\sim1.15$.  A principal difference between using
 our model \xco \ versus the bimodal \xco \ in calculating the
 $\Sigma_{\rm SFR}-\Sigma_{\rm mol}/t_{\rm dyn}$ relation is a
 reduction of scatter.  When measuring the scatter near SFR surface
 density of 1 \msun yr$^{-1}$ kpc$^{-2}$, we find that using our model
 \xco \ versus the bimodal value reduces the scatter by a factor of
 $\sim 5$.

The fact that our model for the $\Sigma_{\rm SFR} - \Sigma_{\rm
  mol}/t_{\rm dyn}$ relation is consistent with the observed one
(though with reduced scatter) is not surprising.  \citet{dad10b} and
\citet{gen10} assume a Milky Way-like \xco \ for their disc galaxies,
and roughly a factor of 5 lower for their mergers. In our model, the
assumption of a bimodal \xco \ for discs and mergers is correct on
average. The mean value for \xco \ for high-\z \ SMGs and low-\z
\ mergers is in fact lower than the mean value for local discs
(c.f. \S~\ref{section:constraints}).  However, many galaxies lie in
the overlap region.  Some local ULIRGs have $X$-factors comparable to
the Galactic average, and some high-\z \ discs have $X$-factors more
similar to the canonical literature ``merger value''.  By modeling the
continuous nature of \xco \ and more properly treating these
intermediate cases, a reduction in scatter is natural.

Finally, we consider $\Sigma_{\rm SFR}$ as a function of $\Sigma_{\rm
  mol}/t_{\rm ff}$, where $t_{\rm ff}$ is the free-fall time within
the dense molecular star-forming clouds in the galaxy.  Both
observations and theory have suggested that the star formation
efficiency per free-fall time $\epsilon_{\rm ff}$ in molecular gas is
approximately constant \citep{kru05,kru07b,eva09,ost11,kru11c}. We
infer $t_{\rm ff}$ from the observable properties of the galaxy
\citep[using the approximations of ][]{kru11c}, together with either a
bimodal \xco\ or our favoured continuous \xco. Krumholz et al. show
that using $t_{\rm ff}$ rather than $t_{\rm dyn}$ makes it possible to
fit the unresolved extragalactic observations, the resolved
observations of Local Group galaxies from \citet{big08}, and
individual molecular clouds in the Milky Way on a single relation, as
illustrated in the left panel of Figure~\ref{figure:tff_ks}.  The
right panel of the Figure shows that this remains true for our
favoured \xco\, and that a star formation law $\Sigma_{\rm SFR} =
\epsilon_{\rm ff} \Sigma_{\rm mol}/t_{\rm ff}$ with $\epsilon_{\rm ff}
\approx 0.01$ remains a good fit to the observational data. As with
the $\Sigma_{\rm SFR} - \Sigma_{\rm mol}/t_{\rm dyn}$ relation, using
our continuous \xco\ actually reduces the scatter, and for the same
reason: our favoured \xco\ produces essentially the same result as the
traditional bimodal \xco\ for galaxies at the extremes of the disc and
merger sequences, but makes the behaviour of \xco\ continuous rather
than discontinuous for the large number of galaxies in the overlap
region.

The results from Figure~\ref{figure:tff_ks} are compatible with the model of
\citet{ost11}. For a disc in vertical hydrostatic equilibrium with
gravity dominated by the gas, and vertical velocity dispersion,
$v_{\rm z}$,
\begin{equation}
\Sigma_{\rm SFR} = \epsilon_{\rm ff} \Sigma_{\rm SFR}/t_{\rm ff} =
\epsilon_{\rm ff} \times 4{\rm G}\Sigma_{\rm mol}^2/(\sqrt3v_{\rm z})
\end{equation}
\citep[][Equation 21]{ost11}.  Comparing to the fit obtained in
Equation~\ref{eq:ks} using our continuous \xco \ relation, we find
that $\epsilon_{\rm ff}/v_{\rm z} = 0.001 (\kms)^{-1} (\Sigma_{\rm
  mol}/100 \ \msun {\rm pc}^{-2})^{-0.05}$.  If $v_{\rm z} \approx 10
\ \kms$ on small scales in the dense neutral gas, as suggested by
\citet{ost11}, this implies $\epsilon_{\rm ff} \approx 0.01$, the
value proposed by \citet{kru05} and \citet{kru07b} and found in
Figure~\ref{figure:tff_ks}.  In the self-regulation theory of
\citet{ost11}, $v_{\rm z}/\epsilon_{\rm ff} \sim (1/3) f_{\rm p}
\times p_*/m_*$, so that $v_{\rm z}$ and $\epsilon_{\rm ff}$ vary
together for a given momentum feedback level.

\subsection{Observational Constraints on the Model and \xco \ Values for Observed Galaxies}
\label{section:constraints}

In order to assess the validity of our parameterisation of \xco
\ (Equation~\ref{eq:xcozsd_ico_formal}), it is worth comparing our
models to the existing observational constraints in the literature.

As discussed in \S~\ref{section:introduction}, galaxy mergers at
low redshift appear to have a range of \xco \ values, from roughly an
order of magnitude below the Galactic mean to comparable to the
Milky-Way average \citep{sol97,dow98,bry99}, though on average the
$X$-factor from local ULIRGs is observed to be below the Galactic mean
\citep{tac08}.  At higher redshifts, the constraints on \xco \ from
inferred mergers (typically submillimetre-selected galaxies) come from
either dynamical mass modeling \citep[][]{tac08}, or dust to gas
ratio arguments \citep{mag11}. The inferred $X$-factors from high-\z
\ SMGs also appear to be lower than the Galactic mean by a factor of
$\sim5$.

There are relatively fewer constraints on \xco \ from high-\z \ discs.
\citet{dad10a} estimated the dynamical masses for resolved CO
observations of high-\z \ \bzk \ disc galaxies. After subtracting off
the measured stellar and assumed dark matter masses, they were able to
derive an \xco \ factor by relating the remaining (presumably \htwo)
mass to the observed CO luminosity.  This method recovered a mean
$X$-factor $\sim 2 \times10^{20}$ \xcounits. This is consistent with
the calculation of \xco \ via dust to gas ratio arguments for a
different \bzk \ galaxy by \citet{mag11}.

In order to investigate how our inferred $X$-factors for observed
galaxies (utilising our model fit) compare to these determinations, on
the left side of Figure~\ref{figure:bzkulirg}, we apply
Equation~\ref{eq:xcozsd_ico_formal} to the observed data from
Figure~\ref{figure:ks}, and plot the derived \xco \ for observed
galaxies, binning separately for local ULIRGs, inferred \zsim 2 discs,
and inferred \zsim 2 mergers.  First, as a consistency check, we
examine the inferred $X$-factors employing
Equation~\ref{eq:xcozsd_ico_formal} from local disc observations, and
denote this by the green dash-dot line in
Figure~\ref{figure:bzkulirg}.  As expected, the derived $X$-factors
from local discs form a relatively tight distribution around the
Galactic mean value of $2-4 \times 10^{20} \xcounits$.

 From Figure~\ref{figure:bzkulirg}, it is evident that observed SMGs
 have extremely low $X$-factors, with the bulk of them a factor of a
 few lower than the Galactic mean value.  Observed ULIRGs show a large
 population of galaxies with lower \xco \ values, with a few
 approaching the Galactic mean.  This is reasonably consistent with
 the range of values reported by \citet{sol97,dow98} and
 \citet{bry99}.

As mentioned, there are far fewer constraints on \xco \ from high-\z
\ discs, with the only constraints placing the $X$-factors near the
Galactic mean value.  The values for some of the observed galaxies in
Figure~\ref{figure:bzkulirg} are consistent with these determinations.
This said, there is a peak in our inferred $X$-factors for high-\z
\ discs at values in-between present-epoch ULIRG $X$-factors and the
Galactic mean value.  Our models therefore predict that attempts to
derive \xco \ for a larger sample of \zsim 2 disc galaxies will indeed
identify some that have $X$-factors more comparable to local ULIRGs.
This means that the true \htwo \ gas masses from observations of
high-\z \ discs and \bzk \ galaxies may be lower than the values
reported in the literature, though by factors of no more than $\sim
2$, depending on the assumed $X$-factor.

Finally, on the right side, we plot the CO line intensity, $W_{\rm
  CO}$ versus both the original assumed \xco \ values for these
observed galaxies, as well as our derived values.  The purpose of this
plot is to show how different galaxy populations are moving between
the centre and right panels of Figure~\ref{figure:ks}.  Again, by and
large, the mergers at present epoch and \zsim 2 tend to keep derived
$X$-factors lower than the Galactic mean, though there are some that
fall to intermediate values.

\section{Discussion of Our Model in the Context of Other Models and Empirical Results}
\label{section:discussion}
A number of groups have begun to address problems related to the
CO-\htwo \ conversion factor in a variety of astrophysical
environments.  As such, in order to build a more complete theoretical
picture of \xco, it is worthwhile understanding how our models fit in
with respect to these models.  The methodologies vary from fully
numerical to empirical.

\citet{glo10} and \citet{glo11} developed magnetohydrodynamic models
of GMC evolution with time-dependent chemistry to follow the formation
and destruction of both \htwo \ and CO.  These models showed that
\htwo \ can survive even in the presence of a photodissociating
radiation field at relatively low column densities owing to
self-shielding, whereas CO can be destroyed more easily.

These models were expanded by \citet{she11a} and \citet{she11b} who
coupled the simulations of \citet{glo11} to large velocity gradient
radiative transfer calculations.  Shetty et al. found that model GMCs with
mean densities, temperatures and metallicities comparable to Galactic
clouds result in $X$-factors comparable to the Galactic mean
value of $\sim2-4 \times 10^{20}$ \xcounits.  Similarly, these authours
found that low metallicity GMCs result in $X$-factors larger
than the Galactic mean.  These results are similar to what we have
found in this study, and have a similar physical reasoning: in low
metallicity regions in their model GMCs, background UV photons
photodissociate CO molecules while the \htwo \ self-shields to
survive.

In the large $\Sigma_{\rm H2}$ regime, our \xco \ results and those of
\citet{she11a,she11b} differ.  Along lines of sight with high-surface
density, \citet{she11b} show that, due to line saturation, \xco
\ begins to rise (see their Figure 2).  While there is quite a bit of
dispersion between model galaxies, we generically see a decrease in
mean \xco \ at high gas surface densities. The differences in \xco
\ are a consequence of differing underlying models.  \citet{she11b}
model individual GMCs that are resolved on sub-parsec scales, whereas
our models are of entire galaxies of entire GMCs resolved at best to
$\sim 70$ pc.  In our simulations, galaxies with higher average mean
GMC surface densities also typically have larger gas temperatures and
velocity dispersions as was discussed in
\S~\ref{section:surfacedensity}.  The velocity dispersions in the
molecular gas often owe to the turbulence driven in a galaxy merger,
and are larger than would be expected if the line widths were solely
due to the virial velocity of the cloud.  Similarly, increased SFRs in
the high surface density gas drive warmer dust, and consequently gas,
temperatures.  The increased temperatures and velocity dispersion
allow for larger CO line intensities per unit \htwo \ gas mass in this
scenario.  In contrast, surface density variations in the GMCs of
\citet{she11b} do not necessarily correlate with larger velocity
dispersions and temperatures in the same way as in a simulated galaxy
merger.  Thus the divergent results in the high $\Sigma_{\rm H2}$
regime owe to different physical processes being modeled.

More recently, \citet{fel11b} have studied the CO-\htwo
\ conversion factor on galaxy-wide scales by coupling the GMC models
of \citet{glo10} and \citet{glo11} with post-processed cosmological
hydrodynamic simulations of galaxy evolution.  While Feldmann et
al. do not have direct information regarding the velocity dispersion
and line widths (and thus refrain from modeling environments such as
ULIRGs), because they utilise the chemical reaction networks of
\citet{glo10} and \citet{glo11}, they are able to study the effects of
metallicity on the $X$-factor with explicit numerical simulations (rather
than the semi-analytic model fit employed here).  Despite their
somewhat more sophisticated treatment of CO formation, the results
they obtain are essentially identical to ours: in low metallicity disc
galaxies, the $X$-factor scales as $Z'$ to a power between -0.4 and
-0.7, consistent with our best fit parameterisation in
Equation~\ref{eq:xcozsd_ico_formal}.

Beyond this, Feldmann et al. relate \xco \ to $\Sigma_{\rm H2}$, and
find variations with scale.  On large scales, they find little
relationship between \xco \ and $\Sigma_{\rm H2}$, whereas on smaller
scales, \xco \ increases with $\Sigma_{\rm H2}$. The reasons for this
apparent discrepancy between the models of Feldmann et al. and us are
the same in the discussion regarding the models of \citet{she11b}.

Finally, \citet{obr09c} applied a Bayesian analysis to literature observational 
data, aiming to relate the $X$-factor in galaxies to various
observables.  These authours found a similar relationship between \xco
\ and $W_{\rm CO}$ as in this paper.  Their fit power-law index
is $\sim -0.31$.

\section{Summary}
\label{section:summary}

We have examined the effects of galactic environment on the CO-\htwo
\ conversion factor in galaxies by coupling simulations of both
quiescent disc and (merger-induced) starburst galaxy evolution at low
and high redshift with dust and molecular line radiative transfer
calculations.  While quiescent disc galaxies at low-\z \ with
metallicities around solar tend to have mean \xco \ values comparable
to the Galactic mean, we find notable regimes in which \xco \ may
differ significantly from this value.  In particular:

\begin{enumerate}

\item {\bf In low metallicity galaxies}, photodissociation destroys CO
  more easily (due to a lack of dust), whereas the \htwo \ can
  self-shield more effectively.  Hence, the amount of \htwo
  \ represented by CO emission rises, and the mean \xco \ is greater
  than the Milky Way average.

\item {\bf In regions of high surface density}, the gas temperature
  and velocity dispersion tend to be rather large.  The former is due to
  heating of dust by young stars, and energy exchange with gas.  The
  latter to the typical origins of high-surface density galaxies:
  either major mergers, or gravitationally unstable clumps in high-\z
  \ gas-rich discs.  In this regime, the velocity-integrated line
  intensity, $W_{\rm CO}$, rises with respect to $\Sigma_{\rm H2}$,
  and there is a net decrease in the mean \xco \ from the Galactic
  mean value.

\item {\bf At high-redshift}, gas-rich discs may have gravitationally
  unstable clumps that have moderate velocity dispersions, and high
  gas temperatures (owing to elevated star formation rates compared to
  the Milky Way).  These galaxies have $X$-factors ranging from the
  Galactic mean to a factor of a few lower.  Our results for rapidly
  star-forming discs at \zsim 2 show that for a given set of physical
  conditions, gas-rich discs at high-\z \ have comparable $X$-factors
  to mergers at low-\z.  Discs and mergers are not inherently
  different with respect to their $X$-factors.  Rather, current local
  conditions determine \xco \ for a particular galaxy.

\end{enumerate}

These results allow us to develop a fitting formula for \xco \ in
terms of gas metallicity and CO line intensity that varies smoothly
(Equation~\ref{eq:xcozsd_ico_formal}).  Applying this formula to CO
detections of galaxies at both low and high-\z, we find that all
versions of the Kennicutt-Schmidt star formation relation
($\Sigma_{\rm SFR}-\Sigma_{\rm mol}$, $\Sigma_{\rm SFR}-\Sigma_{\rm
  mol}/t_{\rm dyn}$ and $\Sigma_{\rm SFR}-\Sigma_{\rm mol}/t_{\rm ff}$)
may be described as unimodal with 
significantly reduced scatter compared to
literature results that use bimodal \xco \ values for discs and mergers.

The results from this work will enable relatively straightforward
application of an $X$-factor which varies with galactic
environment based simply on two observable parameters (metallicity,
and velocity-integrated CO line intensity).

\section*{Acknowledgements} 
This work benefited from work done and conversations had at the Aspen
Center for Physics. Emanuele Daddi, Robert Feldmann, Reinhard Genzel,
Adam Leroy, Patrik Jonsson, Eric Murphy, Rahul Shetty and Andrew
Skemer provided many helpful conversations and suggestions.  We
additionally thank Reinhard Genzel for providing us with digital
copies of the relevant observational data from \citet{tac08} and
\citet{gen11b}, and Emanuele Daddi for providing us with the data from
\citet{dad10b}.  DN and LH acknowledge support from the NSF via grant
AST-1009452.  MK acknowledges support from: an Alfred P. Sloan
Fellowship; the NSF through grants AST-0807739 and CAREER-0955300; and
NASA through Astrophysics Theory and Fundamental Physics grant
NNX09AK31G and a {\it Chandra} Theoretical Research Program grant. ECO
acknowledges support from the NSF via grant AST-0908185.  The
simulations in this paper were run on the Odyssey cluster, supported
by the Harvard Faculty of Arts and Sciences Research Computing Group.

\begin{appendix}
\section{Details of Numerical Modeling}
\label{section:appendix}
In this section, we detail the methodology governing our hydrodynamic
modeling, calculations regarding the chemical and physical state of
the ISM, and radiative transfer. 

\subsection{SPH Galaxy Evolution Simulations}
\label{section:simmethods}

We run \gadget \ smoothed particle hydrodynamic (SPH) simulations for
both model disc galaxies and galaxy mergers.  The aim of these
calculations is to return the spatial distribution and ages of stars,
as well as the metal content and gas content as a function of location
in the model galaxies.  

\gadget \ is a modified version of the publicly available \gadgettwo \
which employs algorithms for feedback from active galactic nuclei
(AGN), and better load balancing on parallel processors.  For a full
description of the underlying algorithms of \gadgettwo, see
\citet{spr05b}.

The ISM is modeled as multi-phase, with cold clouds embedded in a
hotter phase \citep{mck77} which is practically implemented in the
code via hybrid SPH particles \citep{spr03a}.  The phases exchange
mass via radiative cooling of the hotter phase, and supernova heating
within the cold clouds.  Within these cold clouds, stars form
according to a volumetric \citet{sch59} power-law relation with index
1.5 \citep{ken98a}.  The normalisation of this relation is set to
match the surface-density $\Sigma_{\rm SFR}-\Sigma_{\rm gas}$ relation
as observed in local galaxies by \citet{ken98a,ken98b}.  Experiments
have found that imposing a volumetric \citet{sch59} relation results
in a power-law relation comparable to the observed one for the sorts
of galaxies studied here \citep[][though see \citet{sch08} for cases
  when this may not apply]{mih94b,spr00,cox06b}\footnote{In \citet{nar11b},
  we explored the effects of modifying this SFR power-law index, and
  found our modeled $X$-factors were insensitive to variations in the
  SFR index so long as the index is $>1$.  Beyond this, in
  \citet{nar11b}, we argued that a Schmidt index of unity is unlikely
  to describe the starburst environments modeled here as this choice
  of power-law index does not allow mergers to undergo a starburst.
  Similarly, observational \citep{big08} and theoretical work
  \citep{kru09b,ost11} suggest that the SFR index may be superlinear
  in high-surface density environments.}.

The gas is initialised as primordial, with all metals forming as the
simulation progresses.  A mass fraction of stars (consistent with a
Salpeter IMF) are assumed to die instantly in supernovae, and enrich
the surrounding ISM with metals via an instantaneous recycling
approximation using a yield of 0.02 \citep{spr03a}. Hence, gas has a
non-zero metallicity from the first generation of star formation.  We
initialise in this manner in order to probe a large dynamic range of
metallicities in the simulations for both disc galaxies and mergers.
The typical final metallicities of each simulation examined are $\sim$
solar, and we list these in the Table.

Similarly, supernovae impact the surrounding ISM
via energy deposition.  This pressurisation of the ISM is implemented
via an ``effective'' equation of state \citep{rob04,spr05a}.  In the models
presented here, we assume a relatively modest pressurisation of the
ISM ($q_{\rm EOS} = 0.25$) for the model \z=0 galaxies, and more
extreme ($q_{\rm EOS}=1$) for the \z=2 model galaxies.  While the
tests of \citet{nar11b} show that our results are insensitive to the
choice of equation of state within the \citet{spr05b} formalism, we
are intentional with our choices.  The large pressurisation of the ISM
for the high-redshift models is chosen to prevent runaway
fragmentation in the dense, gas-rich environment (to be discussed
shortly).  Because our simulations are non-cosmological, and do not
include gas accretion from the IGM \citep[nor do they have algorithms
  for the inclusion of hot gas in the halo as in][]{mos11a,mos11b},
extreme star formation in the early phases of a galaxy merger would
deplete the gas supply prior to the galaxy merging.  In this limit,
galaxy mergers at high-redshift would not undergo a starburst.
Consequently, we employ a rather stiff EOS.  We note that while this
reduces the amount of fragmentation in the ISM \citep{spr05b}, large
$\sim$kpc-scale clumps do still form and are dynamically unstable even
in the model disc galaxies at \zsim 2, similar to both observations
\citep[e.g.][]{elm09a,elm09b,elm09c, gen11a}, and recent simulations
with other codes \citep[e.g.][]{bou08,dek08,bou10,cev10,bou11}.

Black holes are included in the simulations as sink particles which
accrete according to a Bondi-Lyttleton-Hoyle parameterisation
\citep{bon44} with a fixed maximum rate corresponding to the Eddington
limit.  AGN feedback is included as thermal energy deposited by the
central black hole.  Specifically, 0.5\% of the accreted mass energy
is reinjected spherically into the surrounding ISM which was chosen to
allow merger remnants to match the present-day \magorrian \ relation
\citep{dim05,spr05a,hop07d,hop08a}.  As discussed in the
Appendix of \citet{nar11b}, the inclusion of AGN feedback does not
substantially impact the modeling of \xco \ from galaxies in our
simulations.

The discs are exponential, and initialised according to the
\citet{mo98} model, and embedded in a live cold dark matter halo with
a \citet{her90} density distribution.  The mergers simply involve
discs constructed in this manner.  The initial baryonic gas fractions
are set to 40\% for the \z=0 simulations, and 80\% for the
high-redshift models.  Because the galaxies consume their gas rapidly,
the bulk of the high-redshift snapshots which are analysed in this
study have gas fractions ranging from $f_g = 0.2-0.6$, comparable to
recent CO measurements of \zsim1-2 galaxies by \citet{dad10a} and
\citet{tac10}.

The halo concentration and virial radius for a halo of a given mass is
motivated by cosmological $N$-body simulations, and scaled to match
the expected redshift-evolution following \citet{bul01} and
\citet{rob06a}.  The gravitational softening lengths are set at 100
\hpc for baryons and 200 \hpc for dark matter.  We simulate a wide
range of galaxy baryonic masses, merger mass ratios and merger orbits
for both $\z=0$ and $\z=2$ \ galaxies\footnote{The high-redshift galaxy
  simulations are actually initialised at z=3.  This was chosen so
  that after the $\sim$Gyr typical for the galaxies to reach final
  coalescence when merging, the epoch would correspond roughly to \zsim
  2.}. In the Table, we summarise the physical parameters for the
galaxy evolution models employed for this study, as well as the
reasoning for the inclusion of that particular model.

We note that recently the reliability of SPH for cosmological modeling
has been called into question by comparisons between simulations done
with \gadget \ and others done with the new moving mesh code AREPO
developed by \citet{spr10}.  In particular, \citet{vog11, sij11,ker11}
and \citet{bau11} have shown that under certain conditions, broadly in
circumstances where gas in very different phases is in close proximity
or relative rapid motion, the results with the two codes can be very
different.  In our galaxy-scale simulations, however, we do not
attempt to resolve the different phases of the ISM directly, and
instead rely on subresolution models to describe the gas on small
scales, a strategy dating back to early work by \citet{her89}.  For
this reason, galaxy merger simulations done with AREPO yield results
for the star formation history and mass fraction in newly formed stars
that are virtually identical to those obtained with \gadget.
Therefore, we believe that the results described here are robust with
respect to the numerical algorithm used to perform the hydrodynamics.

\begin{table*}
\label{table:ICs}
\centering
\begin{minipage}{100mm}
\caption{Galaxy evolution simulation parameters. Column 1 refers to
  the model name.  Column 2 is the baryonic mass in \msunend.  Column
  3 is the merger mass ratio, with ``N/A'' denoting when a galaxy is
  an isolated disc.  Please note that we bin all mergers in the text
  as 1:1, 1:3 and 1:10, though these are approximate.  Some
  1:3-designated mergers may be in reality closer to 1:4, for
  example. Columns 4-7 refer to the orbit of a merger (with ``N/A''
  again referring to isolated discs).  Column 8 is the initial
  baryonic gas fraction of the galaxy. Column 9 refers to the redshift
  of the simulation, and Column 10 is the final mass-weighted mean
  metallicity of the system in units of solar metallicity. }
\begin{tabular}{@{}cccccccccc@{}}
\hline Model  & $M_{\rm bar}$& Mass Ratio & $\theta_1$&$\phi_1$ & $\theta_2$ & $\phi_2$ &$f_g$&z&$Z'_{\rm f}$\\ 
&\msun&&&&&&&&$Z_\odot$\\ 
 1 &2 &3 &4 &5 &6 &7 &8 &9 &10\\ \hline \hline\\
z3isob6 &$3.8 \times 10^{11}$&N/A&N/A&N/A&N/A&N/A&0.8&3&0.8\\
z3isob5 &$1.0 \times 10^{11}$&N/A&N/A&N/A&N/A&N/A&0.8&3&0.8\\
z3isob4 &$3.5 \times 10^{10}$&N/A&N/A&N/A&N/A&N/A&0.8&3&1.2\\
z3b6e   &$7.6 \times 10^{11}$&1:1&30&60&-30&45&0.8&3&0.9\\
z3b6b5e &$4.8 \times 10^{11}$&1:3&30&60&-30&45&0.8&3&0.4\\
z3b5e   &$2.0 \times 10^{11}$&1:1&30&60&-30&45&0.8&3&0.9\\
z3b5b4e &$1.4 \times 10^{11}$&1:3&30&60&-30&45&0.8&3&0.5\\
z3b5b3e &$1.1 \times 10^{11}$&1:10&30&60&-30&45&0.8&3&0.8\\
z0isod5 &$4.5 \times 10^{11}$&N/A&N/A&N/A&N/A&N/A&0.4&0&1.7\\
z0isod4 &$1.6 \times 10^{11}$&N/A&N/A&N/A&N/A&N/A&0.4&0&0.8\\
z0isod3 &$5.6 \times 10^{10}$&N/A&N/A&N/A&N/A&N/A&0.4&0&0.9\\
z0d5e   &$8.9 \times 10^{11}$&1:1&30&60&-30&45&0.4&0&1.6\\
z0d5d4e &$6.0 \times 10^{11}$&1:3&30&60&-30&45&0.4&0&0.8\\
z0d5d3e &$5.0 \times 10^{11}$&1:10&30&60&-30&45&0.4&0&0.8\\
z0d4e   &$3.1 \times 10^{11}$&1:1&30&60&-30&45&0.4&0&1.4\\
z0d4h   &$3.1 \times 10^{11}$&1:1&0&0&0&0&0.4&0&1.2\\
z0d4i   &$3.1 \times 10^{11}$&1:1&0&0&71&30&0.4&0&1.1\\
z0d4j   &$3.1 \times 10^{11}$&1:1&-109&90&71&90&0.4&0&1\\
z0d4k   &$3.1 \times 10^{11}$&1:1&-109&30&71&-30&0.4&0&1.1\\
z0d4l   &$3.1 \times 10^{11}$&1:1&-109&30&180&0&0.4&0&1.2\\
z0d4n   &$3.1 \times 10^{11}$&1:1&-109&-30&71&30&0.4&0&1.2\\
z0d4o   &$3.1 \times 10^{11}$&1:1&-109&30&71&-30&0.4&0&1.1\\
z0d4p   &$3.1 \times 10^{11}$&1:1&-109&30&180&0&0.4&0&1.2\\
z0d4d3e   &$2.1 \times 10^{11}$&1:3&30&60&-30&45&0.4&0&0.8\\
z0d4d3i   &$2.1 \times 10^{11}$&1:3&0&0&71&30&0.4&0&1\\
z0d4d3j   &$2.1 \times 10^{11}$&1:3&-109&90&71&90&0.4&0&1.1\\
z0d4d3k   &$2.1 \times 10^{11}$&1:3&-109&30&71&-30&0.4&0&0.9\\
z0d4d3l   &$2.1 \times 10^{11}$&1:3&-109&30&180&0&0.4&0&1\\
z0d4d3m   &$2.1 \times 10^{11}$&1:3&0&0&71&90&0.4&0&1.1\\
z0d4d3n   &$2.1 \times 10^{11}$&1:3&-109&-30&71&30&0.4&0&0.9\\
z0d4d3o   &$2.1 \times 10^{11}$&1:3&-109&30&71&-30&0.4&0&1\\
z0d4d3p   &$2.1 \times 10^{11}$&1:3&-109&30&180&0&0.4&0&1\\
z0d4d2e &$1.7 \times 10^{11}$&1:10&30&60&-30&45&0.4&0&0.8\\

\hline
\end{tabular}
\end{minipage}
\end{table*}

\subsection{Dust Radiative Transfer}

We utilise the publicly available \sunrise \ dust radiative transfer
simulation package \citep{jon06a,jon06b,jon10a,jon10b} for two
purposes: to generate the adaptive mesh from the \gadget \ simulations
on which we run the radiative transfer, and to calculate dust
temperatures in the ISM of the model galaxies.  

In constructing the mesh, the physical conditions are projected onto a
$5^3$ base grid spanning 200 kpc.  The cells recursively refine into
$2^3$ sub-cells based on the refinement criteria that the relative
density variation of metals $\sigma_{\rm \rho m}/<\rho_m>$ should be
less than 0.1 and that the $V$-band optical depth across the cell is
less than unity.  The maximum refinement level is 11, so that the
smallest cells on the grid are of order $\sim 70$ pc across.

For the dust radiative transfer, the sources of light are stellar
clusters and the AGN.  The stars emit a \starburst \ spectrum
\citep{lei99,vaz05} with ages and metallicities known from the SPH
simulations.  The AGN emits a luminosity-dependent SED which is based
on observations of unreddened type I quasars \citep{hop07}.  The
normalisation of the input spectrum is set by the total luminosity of
the black hole(s).

The radiation from stars and the AGN traverses the dusty ISM, and is
scattered, absorbed, and re-emitted.  The evolving dust mass is set by
assuming a constant dust to metals ratio comparable to that of the
Galaxy \citep{dwe98,vla98,cal08}.  We use the \citet{wei01} dust grain
model (with $R = 3.15$) as updated by \citet{dra07}.  The dust and
radiation field are assumed to be in radiative equilibrium, and the
dust temperatures are calculated iteratively.

The low-redshift and high-redshift models differ with regards to the
specification of their ISM properties in the \sunrise \ dust radiative
transfer calculations.  Following \citet{nar11b}, for the low-redshift
models, we assume that young ($<10$ Myr) stellar clusters are embedded in
photodissociation regions (PDRs) and HII regions for some fraction of
their lives.  In this case, the \starburst \ spectrum from the
stellar clusters is replaced by SEDs derived from \mappings
\ photoionisation models \citep{gro08}.  The time-averaged covering
fraction of PDRs is a free-parameter, and is set to be roughly 2-3
Myr.  This value is motivated in part by simulations of \citet{jon10a}
who showed that this parameter choice when applied to models similar
to ours produces synthetic SEDs comparable to those observed in the
Spitzer Infrared Nearby Galaxy Survey sample \citep{ken03}. 

For very massive, gas-rich galaxies, during a merger the stellar
densities can become so high that stellar clusters begin to overlap.
At that point it makes more sense to describe the birthclouds that the
clusters live in as surrounding the entire overlapping (super-)stellar
cluster structures that form (sometimes as massive as many $\times
10^8$ \msunend).  The \mappings \ photoionisation models saturate for
cluster masses of these sizes, and thus become inappropriate to use.
Therefore, for the high-redshift calculations, we abandon the
birthcloud model previously described, and assume the cold ISM is a
uniform medium with a volume filling factor of unity.  While in
reality the ISM may be patchy on these scales, without any information
as to what this patchiness may be like on resolution scales smaller
than the SPH smoothing length, we are forced to make the simplest
assumption possible.  That said, this may be a reasonable choice.
Some evidence exists that in local mergers, a molecular ISM with a
volume filling factor may blanket young stars for the majority of
their lives \citep[e.g.][]{dow98,sak99}.  In any case, the choice is
made primarily due to simulation code constraints.  In \citet{nar11b},
we ran test cases examining the effects of both varying birthcloud
clearing time scales, as well as the uniform volume filling fraction
model, and found minimal effects on the derived $X$-factors.

\subsection{The Thermal and Chemical State of the Molecular ISM}

We assume the neutral mass in each cell in the adaptive mesh is locked
into a spherical isothermal cloud at constant density. The surface
density of the cloud is that which the simulation returns, though we
impose a floor surface density of 100 \msun pc$^{-2}$.  This threshold
surface density, comparable to the surface density of most molecular
clouds in the Local Group
\citep[e.g.][]{sol87,ros03,ros05,ros07,bli06,bli07,bol08}, is imposed to
prevent clouds from having artificially low surface densities in large
cells in the adaptive mesh. When clouds exceed this floor value, we
consider them resolved.

We determine the \htwo-HI balance in these clouds utilising the
analytic model of \citet{kru08,kru09a} and \citet{mck10} which treats the
balance between photodissociation of \htwo \ molecules by Lyman-Werner
band photons against the formation of molecules on grains.  The
equilibrium molecular fraction is given by:
\begin{equation}
\label{eq:kmt}
f_{\rm H2} \approx 1 - \frac{3}{4}\frac{s}{1+0.25s}
\end{equation}
for $s<2$ and $f_{\rm H2} = 0$ for $s\geq 2$.  $s = {\rm ln}
(1+0.6\chi + 0.01\chi^2)/(0.6\tau_{\rm c})$, where $\chi =
0.76(1+3.1Z'^{0.365})$, and $\tau_{\rm c} = 0.066\Sigma_{\rm
  cloud}/(\msun {\rm pc^{-2}})\times Z'$.  $Z'$ is the metallicity
divided by the solar metallicity.  A comparison of this method against
fully-time dependent chemical reaction networks by \citet{kru11b}
finds good correspondence between the methods above metallicities
of $Z' \approx 0.01$.  While we explore the effects of metallicity on
the $X$-factor in galaxies in these models, we never consider cases
with metallicities lower than this value.

With the mass and surface density of the cloud known, the volumetric
density is as well.  We scale this density by a factor
$e^{\sigma_\rho^2/2}$ to account for the turbulent compression of gas,
where numerical simulations suggest that
\begin{equation}
\sigma_\rho^2 \approx {\rm ln}(1+3M_{1D}^2/4)
\end{equation}
is a good approximation, where $M_{\rm 1D}$ is the 1 dimensional Mach
number of the turbulence \citep{ost01,pad02,lem08,pri11}. The Mach
number is calculated by assuming the cloud temperature is 10 K (to
avoid having to iterate as the temperature we will calculate is dependent
on the density of the gas).

The velocity dispersion of the gas is determined as the mean square
sum of the subgrid turbulent velocity dispersion and the resolved
nonthermal velocity dispersion.  The subgrid dispersion is calculated
from the external pressure, $P = \rho_{\rm cell}\sigma^2$, and has an
imposed ceiling of 10 \kmsend, as suggested from galactic simulations which
model turbulent energy driving and dissipation
\citep{dib06,jou09,ost11}.  The resolved nonthermal component is
calculated from the standard deviation in the velocity dispersion in
the nearest neighbour cells in the $\hat{x}, \hat{y}$ and $\hat{z}$
directions.  In cases where the cloud is unresolved, we simply assume
the GMC is in virial balance, with virial parameter $\alpha_{\rm vir}
= 5\sigma^2_{\rm vir}R/(GM)$ of order unity so that
\begin{equation}
\sigma_{\rm vir} = 2.2 \ \kms \  \left[\frac{M}{10^5 \ \msun}\right]^{1/4}
\end{equation}
where $M$ is the mass of the cloud.

Finally, we calculate the temperature of the GMC based on the model
developed by \citet{kru11a}.  In this, we calculate the heating and
cooling processes on the gas, the heating and cooling of the dust, and
the energy exchange between the two.  The heating processes of the gas
are the grain photoelectric effect, and cosmic rays, and the cooling
occurs via either CO or CII line emission.  The dust is heated by the
background radiation field, and cools thermally.  Formally, if we
denote heating processes by $\Gamma$, cooling by $\Lambda$, and energy
exchange via $\Psi$, then we have:

\begin{eqnarray}
\Gamma_{\rm pe} + \Gamma_{\rm CR} - \Lambda_{\rm line} + \Psi_{\rm gd} = 0\\
\Gamma_{\rm dust} - \Lambda_{\rm dust} - \Psi_{\rm gd} = 0  
\end{eqnarray}
For brevity, we refer the reader to \citet{kru11a} and \citet{nar11b} for the equations
regarding the photoelectric effect and cosmic ray heating terms.

The gas cooling is assumed to happen either via CII or CO line
emission.  Following \citet{wol10}, we approximate the fraction of
hydrogen where CO is the dominant form of carbon by:
\begin{equation}
\label{eq:abundance}
f_{\rm CO} = f_{\rm H2} \times e^{-4(0.53-0.045 {\rm
    ln}\frac{G_0'}{n_{\rm H}/{\rm cm^{-3}}}-0.097 {\rm ln}Z')/A_{\rm v}}
\end{equation}
which is consistent with the numerical work of \citet{glo11}.  When
$f_{\rm CO}$ is above 50\%, we assume the cooling happens via CO line
cooling.  Otherwise, CII is the dominant coolant.   The
extinction is converted via:
\begin{equation}
A_{\rm V} = \frac{N_{\rm H}}{1.87 \times 10^{21} {\rm cm^{-2}}}Z'
\end{equation}
\citep[e.g.][]{wat11}. We assume the dust is heated primarily by
background infrared radiation as heating by UV radiation is likely to
be highly suppressed by extinction.  The IR radiation field is known
from the \sunrise \ dust radiative transfer modeling. The escape of
the photons from the GMCs is calculated via the public escape
probability code of \citet{kru07}, which we describe in the subsequent
section along with the other molecular line radiative transfer
equations.

\subsection{Molecular Line Radiative Transfer}

Once we know the physical state of the GMCs in our model galaxies, we
perform molecular line radiative transfer to calculate the CO
intensity from the clouds.  We first utilise an escape probability
formalism to determine the escape fraction of photons from individual
clouds, and then 3D non-LTE Monte Carlo molecular line radiative
transfer calculations in order to calculate the transport of photons
through the galaxies.

The escape probabilities are calculated using the publicly
available code described in \citet{kru07}.  The levels are assumed to
be in statistical equilibrium, and are calculated by balancing the
excitation and deexcitation of CO by collisions with \htwo \ and He,
absorption, stimulated emission and spontaneous emission via the rate
equations:
\begin{eqnarray}
\label{eq:kmt_stateq}
\sum_l(C_{lu} + \beta_{lu}A_{lu})f_l =
\left[\sum_u(C_{ul}+\beta_{ul}A_{ul})\right]f_u\\
\sum_if_i = 1
\end{eqnarray}
where $C$ are the collisional rates, $f$ the fractional level
populations, $A$ are the Einstein rate coefficients, and $\beta_{ul}$
is the escape probability for transition $u\rightarrow l$.  The rate
equations are re-arranged as an eigenvalue problem and solved with
public packages available in the GNU Scientific Library.

$\beta$ is well-fit by the approximate relation \citep{dra11} 
\begin{equation}
\label{eq:kmt_beta}
\beta_{ul} \approx \frac{1}{1+0.5\tau_{ul}}
\end{equation}
where the optical depth is 
\begin{equation}
\label{eq:kmt_tau}
\tau_{ul} =
\frac{g_u}{g_l}\frac{3A_{ul}\lambda_{ul}^3}{16(2\pi)^{3/2}\sigma}QN_{\rm
  H2}f_l\left(1-\frac{f_ug_l}{f_lg_u}\right)
\end{equation}
where $Q$ is the abundance of CO with respect to \htwo, $g_l$ and
$g_u$ are the statistical weights of the levels, $N_{\rm H2}$ is the
column density of \htwo \ through the cloud, $\lambda_{ul}$ is the
wavelength of the transition, and $\sigma$ is the velocity dispersion
in the cloud.  We iterate
Equations~\ref{eq:kmt_stateq}-\ref{eq:kmt_tau} with standard
Newton-Raphson methods until $\beta$ and the populations are known for
all levels simultaneously.

We then propagate the photons which escape the model GMCs through the
galaxy following the methods of \citet{ber79}\footnote{In
  \citet{nar11b}, we fully describe the equations employed for the
  galaxy-wide radiative transfer.  The interested reader should refer
  to this paper for more detail, and here we summarise the relevant
  physical processes for the sake of brevity.}.  The intrinsic line
profile function is assumed to be Gaussian in nature with width given
by the Doppler-width.  The model photons are emitted isotropically,
and can be absorbed by GMCs they encounter along the path with a
probability determined both by the volume filling factor of the GMC in
the cell, as well as the absorption line profile in the cell (thus
accounting for decreasing effective optical depths owing to the
absorption profile and emission profile shifting out of resonance when
the line of sight velocity difference between the emitting and
absorbing cell is large).  When all the model photons have been
emitted, the rate equations are again evaluated, level populations
updated, and new model photons are emitted.  This process is iterated
upon until convergence (a fractional difference in the level
populations of $1\times10^{-3}$) is reached.  In practice, this
happens relatively quickly as velocity gradients in the galaxy render
the ISM globally optically thin to most photons.

\section{Simulated Observational Properties of Model Galaxies}
\label{section:simulatedobservations}

Because our methodology involves dust radiative transfer, we are able
to make some comparisons to observed properties of galaxies.  Here,
we present a few comparisons to observations primarily to demonstrate
that our simulated galaxies serve as reasonable analogs to the sorts
of galaxies typically observed in large CO surveys.

\begin{figure}
\hspace{-1cm}
\includegraphics[angle=90,scale=0.4]{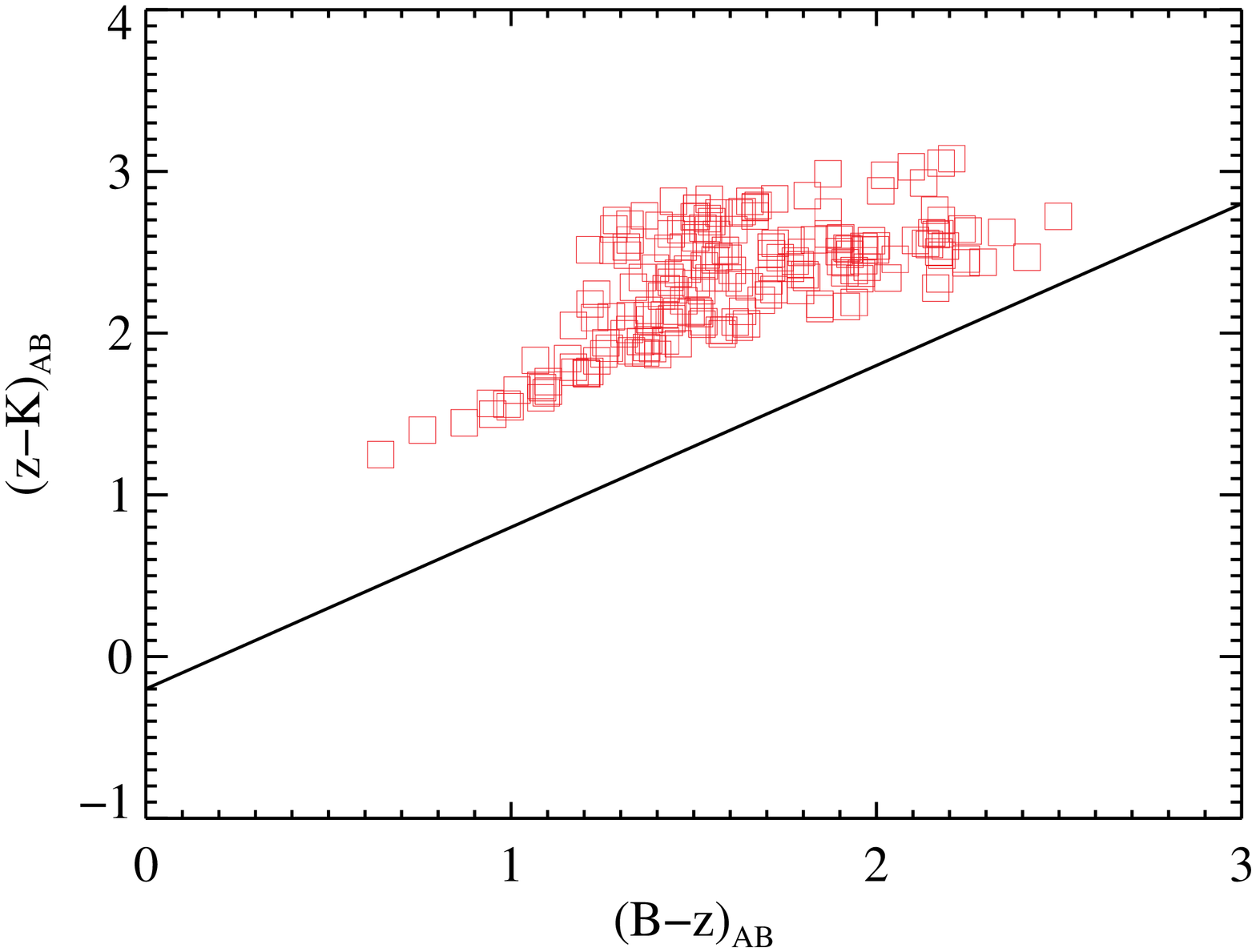}
\caption{Synthetic $(B-z)$ and $(z-K)$ colours of model high-\z \ disc
  galaxies.  Each red square represents an individual snapshot from
  models z3isob6, z3isob5 and z3isob4. The model galaxies are
  redshifted to \z \ =  2, and the colours are observed-frame.
  Typically, galaxies above the solid line are observed to be
  star-forming discs \citep{dad04}, and are representative of the
  sorts of ``normal'' (not undergoing a burst of star formation)
  high-\z \ galaxies observed in CO.  Our model discs have optical
  observed-frame colours comparable to high-\z
  \ discs.  \label{figure:bzk}}
\end{figure}

At high-redshift, the bulk of the galaxies observed in CO are
identified according to 
two selection techniques: the \bzk \ population, and
submillimetre-selected galaxies.  \bzk \ galaxies are defined by their
optical colour ratios, and galaxies with blue $B-z$ colours, and red
$z-K$ colours are typically denoted as star-forming \bzk \ galaxies
due to the presence of a Balmer break.  In Figure~\ref{figure:bzk}, we
examine the optical colours of our model high-\z \ discs.  We redshift
the discs to \z=2, and plot the observed $(B-z)$ and $(z-K)$ colours
as a comparison to \zsim 2 \bzk \ galaxies. As originally discussed by
\citet{dad04}, galaxies which fall above the line denoted in
Figure~\ref{figure:bzk} tend to be
disc-like in morphology \citep{dad05,for09}.  As is evident, our three
model high-\z \ discs all fall in the same space as observed \bzk
\ galaxies.

\citet{nar10a} and \citet{hay11} presented a merger-driven model for
the formation of high-\z \ submillimetre galaxies utilising
merger-models similar to our high-redshift mergers.  These galaxies
were seen to match the mean SEDs of observed SMGs, as well as the
typical CO excitation patterns, line widths, and morphologies
\citep{nar09}.  Furthermore, when convolving submillimetre-luminous
duty cycles with theoretical galaxy merger rates and observed galaxy
mass functions, \citet{hay10} and C. Hayward et al. (in prep.) find
that this merger-driven model provides a reasonable match to the
observed number counts and redshift distribution of SMGs.  In this
sense, our model high-\z \ mergers are likely reasonable analogs for
the sorts of high-\z \ starbursts observed in CO
\citep[e.g.][]{gre05,tac06,bou07,tac08,bot09,gen10}.

\begin{figure}
\hspace{-1cm}
\includegraphics[angle=90,scale=0.45]{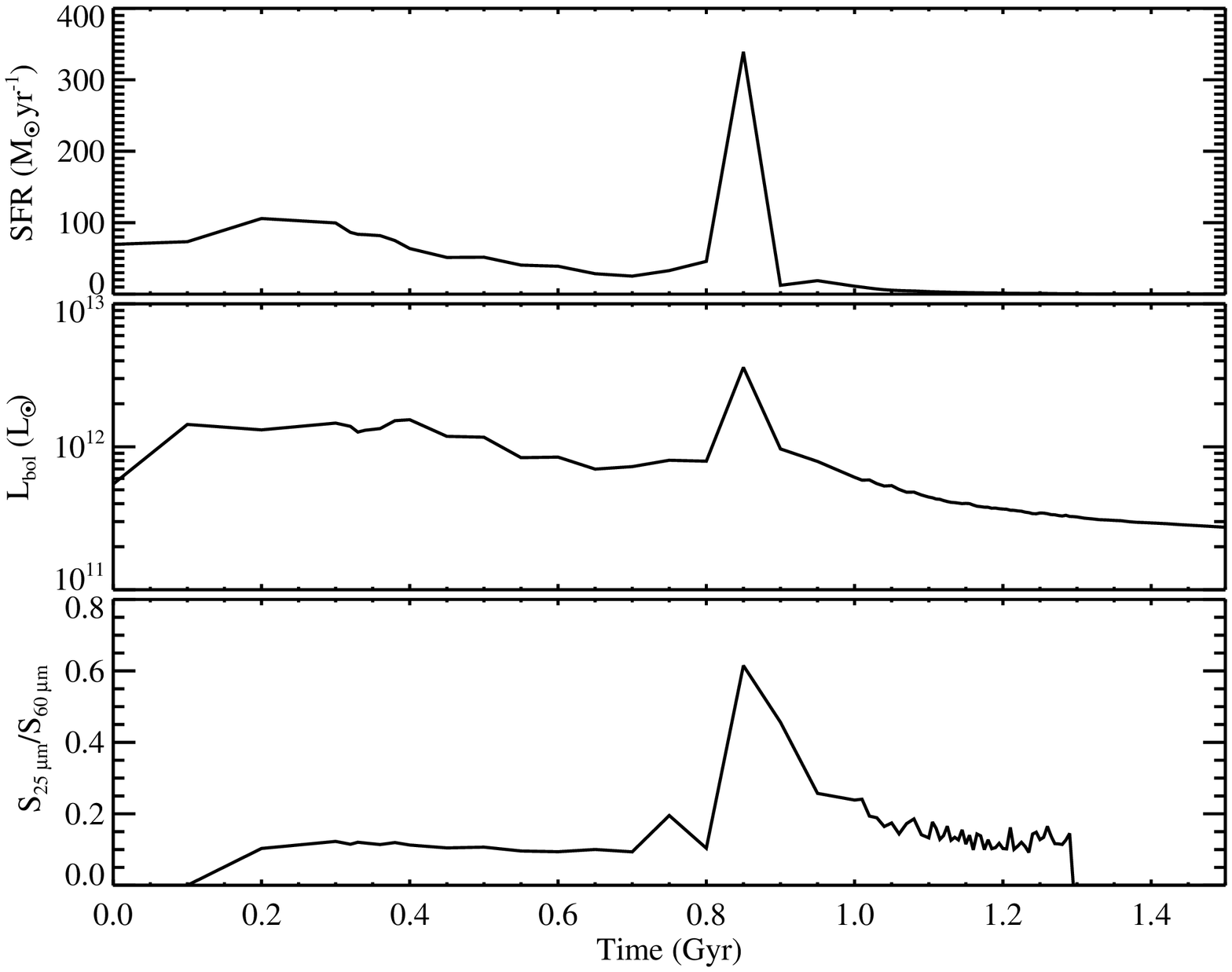}
\caption{Evolution of SFR, \lbol, and IRAS colour ratios for an
  example local merger (model z0d4e).  Upon merging, the model galaxy
  undergoes a vigorous starburst, is selectable as a ULIRG, and
  exhibits IRAS $25/60 \micron$ ratios comparable to many local
  ULIRGs.  Therefore, our simulated mergers 
  are reasonable analogs for observed
  ULIRGs in the local Universe. \label{figure:sfr_lir}}
\end{figure}

Our simulated mergers at low-redshift are similarly reasonable models
for present-epoch ULIRGs.  In Figure~\ref{figure:sfr_lir}, we show the
evolution of the star formation rate, bolometric luminosity, and IRAS
$25\micron/60\micron$ flux density ratios for merger model z0d4e, the
fiducial merger from \citet{nar11b}.  As is evident, when the galaxies
merge (around $T\approx 0.85$ Gyr), the SFR undergoes a burst of
comparable magnitude to the most heavily star forming galaxies
locally.  Similarly, the bolometric luminosity rises to the point that
the galaxy would be selectable as a ULIRG.  The IRAS infrared colours
transition from ``cool'' to ``warm'' as the starburst and AGN heat the
gas \citep{you09}.

Beyond these few examples, a number of other comparisons to
observations exist in the literature for this exact same set of
simulations.  For the high-redshift sample, the galaxies have been
shown to reproduce observed properties of Spitzer-selected $24
\micron$ sources \citep{nar10b} and properties of bright quasars
\citep{hop06,hop06b}.  The low-redshift galaxy mergers have been shown
to reproduce kinematic properties of early-type galaxies
\citep{cox06b}, observed $X$-ray properties of and metallicities in
mergers \citep{cox06c,tor11}, the broad-band colours of post-starburst
galaxies \citep{sny11}, and the structural properties of merging and
elliptical galaxies \citep{hop08b,hop08c,hop08e,hop09}.

\section{Enhanced Cosmic Ray Ionisation Rates in Starbursts}
\label{section:cosmicrays}
In this section, we assess the validity of our assumption of a
Galactic cosmic ray ionisation rate.  Cosmic rays may affect \xco \ in
galaxies by providing a source of heat for the molecular gas.
Recalling the methodology presented in this Appendix, we assume a
Galactic cosmic ray ionisation rate in all model galaxies.  It is
conceivable that this assumption may break down in
starbursts. Observations of M82 by the VERITAS collaboration suggest a
link between star formation and cosmic rays \citep{acc09}. Similarly,
detections of $\gamma$-rays from the Galaxy, LMC, NGC 253 and M82 by
the FERMI group find a reasonable correlation between the $\gamma$-ray
luminosity, and the product of the supernovae rate and gas mass
\citep{abd10b}.  Since $\gamma$-rays are the product of cosmic rays
with hydrogen atoms, a reasonable scaling may be that the cosmic ray
ionisation rate is linearly related to the star formation rate.
Indeed, observations of local starbursts have suggested a potential
for cosmic-ray heated gas \citep[e.g.][]{hai08}.

In order to test this, we have run a series of simulations in which we
increase the cosmic ray ionisation rate linearly with SFR.  We assume
a SFR of 2 \msunyr for the Milky Way \citep{rob10}.  We span the
range of physical conditions in our models, running a high-\z \ 1:1
merger, high-\z \ disc, low \z \ 1:1 merger, and low-\z \ disc
simulation.

In these examples, an increased cosmic ray ionisation rate which scales
linearly with the SFR does little to change \xco.  We look at a
limiting case to illustrate why this is.  In merger-induced ULIRGs,
where the SFR/cosmic ray flux can be a factor of as much as $200$
higher than the Galactic value (e.g. Submillimetre galaxies), the
minimum gas temperature imposed by cosmic ray-heating is $\sim 30-50$
K, depending on the density of the gas \citep{pap10b}.  In these
environments, the dust temperature is typically not very different
from this.  For example, in \citet{nar11b}, we found the mass-weighted
mean dust temperature of a local merger was $\sim 70$ K.  In short,
the scenarios in which the cosmic ray ionisation rate may be increased
have a large gas temperature anyway due to high dust temperatures and
efficient energy exchange between the two at high densities.  We
therefore 
tentatively conclude that enhanced cosmic ray ionisation rates do not
affect \xco \ in starburst environments, though we caution that if the
cosmic ray heating rate rises much faster than linearly with SFR
\citep[see, for example, ][]{bra03}, then this inference may break
down.  Based on the calculations of \citet{pap10b}, we estimate that
the cosmic ray heating would affect our results significantly only if
the ratios of the cosmic ray heating rates in mergers and isolated
galaxies were at least 10 times greater than their ratios of star
formation rate.

\end{appendix}


\begin{thebibliography}{189}
\expandafter\ifx\csname natexlab\endcsname\relax\def\natexlab#1{#1}\fi

\bibitem[{{Abdo} {et~al.}(2010{\natexlab{a}})}]{abd10b}
{Abdo}, A.~A. {et~al.} 2010{\natexlab{a}}, \apjl, 709, L152

\bibitem[{{Abdo} {et~al.}(2010{\natexlab{b}})}]{abd10}
---. 2010{\natexlab{b}}, \apj, 710, 133

\bibitem[{{Acciari} {et~al.}(2009)}]{acc09}
{Acciari}, V.~A. {et~al.} 2009, \nat, 462, 770

\bibitem[{{Arimoto} {et~al.}(1996){Arimoto}, {Sofue}, \& {Tsujimoto}}]{ari96}
{Arimoto}, N., {Sofue}, Y., \& {Tsujimoto}, T. 1996, PASJ, 48, 275

\bibitem[{{Barnes} \& {Hernquist}(1996)}]{bar96}
{Barnes}, J.~E. \& {Hernquist}, L. 1996, \apj, 471, 115

\bibitem[{{Barnes} \& {Hernquist}(1991)}]{bar91}
{Barnes}, J.~E. \& {Hernquist}, L.~E. 1991, \apjl, 370, L65

\bibitem[{{Bauer} \& {Springel}(2011)}]{bau11}
{Bauer}, A. \& {Springel}, V. 2011, arXiv/1109.4413

\bibitem[{{Bell} {et~al.}(2006){Bell}, {Roueff}, {Viti}, \& {Williams}}]{bel06}
{Bell}, T.~A., {Roueff}, E., {Viti}, S., \& {Williams}, D.~A. 2006, \mnras,
  371, 1865

\bibitem[{{Bell} {et~al.}(2007){Bell}, {Viti}, \& {Williams}}]{bel07}
{Bell}, T.~A., {Viti}, S., \& {Williams}, D.~A. 2007, \mnras, 378, 983

\bibitem[{{Bernes}(1979)}]{ber79}
{Bernes}, C. 1979, \aap, 73, 67

\bibitem[{{Bertsch} {et~al.}(1993){Bertsch}, {Dame}, {Fichtel}, {Hunter},
  {Sreekumar}, {Stacy}, \& {Thaddeus}}]{ber93}
{Bertsch}, D.~L., {Dame}, T.~M., {Fichtel}, C.~E., {Hunter}, S.~D.,
  {Sreekumar}, P., {Stacy}, J.~G., \& {Thaddeus}, P. 1993, \apj, 416, 587

\bibitem[{{Bigiel} {et~al.}(2008){Bigiel}, {Leroy}, {Walter}, {Brinks}, {de
  Blok}, {Madore}, \& {Thornley}}]{big08}
{Bigiel}, F., {Leroy}, A., {Walter}, F., {Brinks}, E., {de Blok}, W.~J.~G.,
  {Madore}, B., \& {Thornley}, M.~D. 2008, \aj, 136, 2846

\bibitem[{{Blitz} {et~al.}(2007){Blitz}, {Fukui}, {Kawamura}, {Leroy},
  {Mizuno}, \& {Rosolowsky}}]{bli07}
{Blitz}, L., {Fukui}, Y., {Kawamura}, A., {Leroy}, A., {Mizuno}, N., \&
  {Rosolowsky}, E. 2007, in Protostars and Planets V, ed. B.~{Reipurth},
  D.~{Jewitt}, \& K.~{Keil}, 81--96

\bibitem[{{Blitz} \& {Rosolowsky}(2006)}]{bli06}
{Blitz}, L. \& {Rosolowsky}, E. 2006, \apj, 650, 933

\bibitem[{{Bloemen} {et~al.}(1986){Bloemen}, {Strong}, {Mayer-Hasselwander},
  {Blitz}, {Cohen}, {Dame}, {Grabelsky}, {Thaddeus}, {Hermsen}, \&
  {Lebrun}}]{blo86}
{Bloemen}, J.~B.~G.~M., {Strong}, A.~W., {Mayer-Hasselwander}, H.~A., {Blitz},
  L., {Cohen}, R.~S., {Dame}, T.~M., {Grabelsky}, D.~A., {Thaddeus}, P.,
  {Hermsen}, W., \& {Lebrun}, F. 1986, \aap, 154, 25

\bibitem[{{Bolatto} {et~al.}(2008){Bolatto}, {Leroy}, {Rosolowsky}, {Walter},
  \& {Blitz}}]{bol08}
{Bolatto}, A.~D., {Leroy}, A.~K., {Rosolowsky}, E., {Walter}, F., \& {Blitz},
  L. 2008, \apj, 686, 948

\bibitem[{{Bondi} \& {Hoyle}(1944)}]{bon44}
{Bondi}, H. \& {Hoyle}, F. 1944, \mnras, 104, 273

\bibitem[{{Boselli} {et~al.}(2002){Boselli}, {Lequeux}, \& {Gavazzi}}]{bos02}
{Boselli}, A., {Lequeux}, J., \& {Gavazzi}, G. 2002, AP\&SS, 281, 127

\bibitem[{{Bothwell} {et~al.}(2010)}]{bot09}
{Bothwell}, M.~S. {et~al.} 2010, \mnras, 405, 219

\bibitem[{{Bouch{\'e}} {et~al.}(2007)}]{bou07}
{Bouch{\'e}}, N. {et~al.} 2007, \apj, 671, 303

\bibitem[{{Bournaud} {et~al.}(2011){Bournaud}, {Chapon}, {Teyssier}, {Powell},
  {Elmegreen}, {Elmegreen}, {Duc}, {Contini}, {Epinat}, \& {Shapiro}}]{bou11}
{Bournaud}, F., {Chapon}, D., {Teyssier}, R., {Powell}, L.~C., {Elmegreen},
  B.~G., {Elmegreen}, D.~M., {Duc}, P., {Contini}, T., {Epinat}, B., \&
  {Shapiro}, K.~L. 2011, \apj, 730, 4

\bibitem[{{Bournaud} {et~al.}(2008){Bournaud}, {Daddi}, {Elmegreen},
  {Elmegreen}, {Nesvadba}, {Vanzella}, {Di Matteo}, {Le Tiran}, {Lehnert}, \&
  {Elbaz}}]{bou08}
{Bournaud}, F., {Daddi}, E., {Elmegreen}, B.~G., {Elmegreen}, D.~M.,
  {Nesvadba}, N., {Vanzella}, E., {Di Matteo}, P., {Le Tiran}, L., {Lehnert},
  M., \& {Elbaz}, D. 2008, \aap, 486, 741

\bibitem[{{Bournaud} {et~al.}(2010){Bournaud}, {Elmegreen}, {Teyssier},
  {Block}, \& {Puerari}}]{bou10}
{Bournaud}, F., {Elmegreen}, B.~G., {Teyssier}, R., {Block}, D.~L., \&
  {Puerari}, I. 2010, \mnras, 409, 1088

\bibitem[{{Bradford} {et~al.}(2003){Bradford}, {Nikola}, {Stacey}, {Bolatto},
  {Jackson}, {Savage}, {Davidson}, \& {Higdon}}]{bra03}
{Bradford}, C.~M., {Nikola}, T., {Stacey}, G.~J., {Bolatto}, A.~D., {Jackson},
  J.~M., {Savage}, M.~L., {Davidson}, J.~A., \& {Higdon}, S.~J. 2003, \apj,
  586, 891

\bibitem[{{Bryant} \& {Scoville}(1999)}]{bry99}
{Bryant}, P.~M. \& {Scoville}, N.~Z. 1999, \aj, 117, 2632

\bibitem[{{Bullock} {et~al.}(2001){Bullock}, {Kolatt}, {Sigad}, {Somerville},
  {Kravtsov}, {Klypin}, {Primack}, \& {Dekel}}]{bul01}
{Bullock}, J.~S., {Kolatt}, T.~S., {Sigad}, Y., {Somerville}, R.~S.,
  {Kravtsov}, A.~V., {Klypin}, A.~A., {Primack}, J.~R., \& {Dekel}, A. 2001,
  \mnras, 321, 559

\bibitem[{{Calura} {et~al.}(2008){Calura}, {Pipino}, \& {Matteucci}}]{cal08}
{Calura}, F., {Pipino}, A., \& {Matteucci}, F. 2008, \aap, 479, 669

\bibitem[{{Carilli} {et~al.}(2010)}]{car10}
{Carilli}, C.~L. {et~al.} 2010, \apj, 714, 1407

\bibitem[{{Ceverino} {et~al.}(2010){Ceverino}, {Dekel}, \& {Bournaud}}]{cev10}
{Ceverino}, D., {Dekel}, A., \& {Bournaud}, F. 2010, \mnras, 404, 2151

\bibitem[{{Cox} {et~al.}(2006{\natexlab{a}}){Cox}, {Di Matteo}, {Hernquist},
  {Hopkins}, {Robertson}, \& {Springel}}]{cox06c}
{Cox}, T.~J., {Di Matteo}, T., {Hernquist}, L., {Hopkins}, P.~F., {Robertson},
  B., \& {Springel}, V. 2006{\natexlab{a}}, \apj, 643, 692

\bibitem[{{Cox} {et~al.}(2006{\natexlab{b}})}]{cox06b}
{Cox}, T.~J. {et~al.} 2006{\natexlab{b}}, \apj, 650, 791

\bibitem[{{Cresci} {et~al.}(2010){Cresci}, {Mannucci}, {Maiolino}, {Marconi},
  {Gnerucci}, \& {Magrini}}]{cre10}
{Cresci}, G., {Mannucci}, F., {Maiolino}, R., {Marconi}, A., {Gnerucci}, A., \&
  {Magrini}, L. 2010, \nat, 467, 811

\bibitem[{{Crosthwaite} \& {Turner}(2007)}]{cro07}
{Crosthwaite}, L.~P. \& {Turner}, J.~L. 2007, \aj, 134, 1827

\bibitem[{{Daddi} {et~al.}(2004)}]{dad04}
{Daddi}, E. {et~al.} 2004, \apj, 617, 746

\bibitem[{{Daddi} {et~al.}(2005)}]{dad05}
---. 2005, \apjl, 631, L13

\bibitem[{{Daddi} {et~al.}(2007)}]{dad07}
---. 2007, \apj, 670, 156

\bibitem[{{Daddi} {et~al.}(2010{\natexlab{a}})}]{dad10b}
---. 2010{\natexlab{a}}, \apjl, 714, L118

\bibitem[{{Daddi} {et~al.}(2010{\natexlab{b}})}]{dad10a}
---. 2010{\natexlab{b}}, \apj, 713, 686

\bibitem[{{Dame} {et~al.}(2001){Dame}, {Hartmann}, \& {Thaddeus}}]{dam01}
{Dame}, T.~M., {Hartmann}, D., \& {Thaddeus}, P. 2001, \apj, 547, 792

\bibitem[{{Dav{\'e}} {et~al.}(2010){Dav{\'e}}, {Finlator}, {Oppenheimer},
  {Fardal}, {Katz}, {Kere{\v s}}, \& {Weinberg}}]{dav10}
{Dav{\'e}}, R., {Finlator}, K., {Oppenheimer}, B.~D., {Fardal}, M., {Katz}, N.,
  {Kere{\v s}}, D., \& {Weinberg}, D.~H. 2010, \mnras, 404, 1355

\bibitem[{{de Vries} {et~al.}(1987){de Vries}, {Thaddeus}, \&
  {Heithausen}}]{dev87}
{de Vries}, H.~W., {Thaddeus}, P., \& {Heithausen}, A. 1987, \apj, 319, 723

\bibitem[{{Dekel} {et~al.}(2009){Dekel}, {Birnboim}, {Engel}, {Freundlich},
  {Goerdt}, {Mumcuoglu}, {Neistein}, {Pichon}, {Teyssier}, \& {Zinger}}]{dek08}
{Dekel}, A., {Birnboim}, Y., {Engel}, G., {Freundlich}, J., {Goerdt}, T.,
  {Mumcuoglu}, M., {Neistein}, E., {Pichon}, C., {Teyssier}, R., \& {Zinger},
  E. 2009, \nat, 457, 451

\bibitem[{{Delahaye} {et~al.}(2011){Delahaye}, {Fiasson}, {Pohl}, \&
  {Salati}}]{del11}
{Delahaye}, T., {Fiasson}, A., {Pohl}, M., \& {Salati}, P. 2011, \aap, 531,
  A37+

\bibitem[{{Di Matteo} {et~al.}(2005){Di Matteo}, {Springel}, \&
  {Hernquist}}]{dim05}
{Di Matteo}, T., {Springel}, V., \& {Hernquist}, L. 2005, \nat, 433, 604

\bibitem[{{Dib} {et~al.}(2006){Dib}, {Bell}, \& {Burkert}}]{dib06}
{Dib}, S., {Bell}, E., \& {Burkert}, A. 2006, \apj, 638, 797

\bibitem[{{Dickman}(1975)}]{dic75}
{Dickman}, R.~L. 1975, \apj, 202, 50

\bibitem[{{Donovan Meyer} {et~al.}(2011){Donovan Meyer}, {Koda}, {Momose},
  {Fukuhara}, {Mooney}, {Towers}, {Egusa}, {Kennicutt}, {Kuno}, {Carty},
  {Sawada}, \& {Scoville}}]{don11}
{Donovan Meyer}, J., {Koda}, J., {Momose}, R., {Fukuhara}, M., {Mooney}, T.,
  {Towers}, S., {Egusa}, F., {Kennicutt}, R., {Kuno}, N., {Carty}, M.,
  {Sawada}, T., \& {Scoville}, N. 2011, arXiv/1109.6272

\bibitem[{{Downes} \& {Solomon}(1998)}]{dow98}
{Downes}, D. \& {Solomon}, P.~M. 1998, \apj, 507, 615

\bibitem[{{Downes} \& {Solomon}(2003)}]{dow03}
---. 2003, \apj, 582, 37

\bibitem[{{Downes} {et~al.}(1993){Downes}, {Solomon}, \& {Radford}}]{dow93}
{Downes}, D., {Solomon}, P.~M., \& {Radford}, S.~J.~E. 1993, \apjl, 414, L13

\bibitem[{{Draine}(2011)}]{dra11}
{Draine}, B.~T. 2011, {Physics of the Interstellar and Intergalactic Medium},
  ed. {Draine, B.~T.}

\bibitem[{{Draine} \& {Li}(2007)}]{dra07}
{Draine}, B.~T. \& {Li}, A. 2007, \apj, 657, 810

\bibitem[{{Dwek}(1998)}]{dwe98}
{Dwek}, E. 1998, \apj, 501, 643

\bibitem[{{Elmegreen} {et~al.}(2008){Elmegreen}, {Bournaud}, \&
  {Elmegreen}}]{elm09c}
{Elmegreen}, B.~G., {Bournaud}, F., \& {Elmegreen}, D.~M. 2008, \apj, 688, 67

\bibitem[{{Elmegreen} {et~al.}(2009{\natexlab{a}}){Elmegreen}, {Elmegreen},
  {Fernandez}, \& {Lemonias}}]{elm09b}
{Elmegreen}, B.~G., {Elmegreen}, D.~M., {Fernandez}, M.~X., \& {Lemonias},
  J.~J. 2009{\natexlab{a}}, \apj, 692, 12

\bibitem[{{Elmegreen} {et~al.}(2009{\natexlab{b}}){Elmegreen}, {Elmegreen},
  {Marcus}, {Shahinyan}, {Yau}, \& {Petersen}}]{elm09a}
{Elmegreen}, D.~M., {Elmegreen}, B.~G., {Marcus}, M.~T., {Shahinyan}, K.,
  {Yau}, A., \& {Petersen}, M. 2009{\natexlab{b}}, \apj, 701, 306

\bibitem[{{Engel} {et~al.}(2010){Engel}, {Tacconi}, {Davies}, {Neri}, {Smail},
  {Chapman}, {Genzel}, {Cox}, {Greve}, {Ivison}, {Blain}, {Bertoldi}, \&
  {Omont}}]{eng10}
{Engel}, H., {Tacconi}, L.~J., {Davies}, R.~I., {Neri}, R., {Smail}, I.,
  {Chapman}, S.~C., {Genzel}, R., {Cox}, P., {Greve}, T.~R., {Ivison}, R.~J.,
  {Blain}, A., {Bertoldi}, F., \& {Omont}, A. 2010, \apj, 724, 233

\bibitem[{{Erb} {et~al.}(2006){Erb}, {Steidel}, {Shapley}, {Pettini}, {Reddy},
  \& {Adelberger}}]{erb06a}
{Erb}, D.~K., {Steidel}, C.~C., {Shapley}, A.~E., {Pettini}, M., {Reddy},
  N.~A., \& {Adelberger}, K.~L. 2006, \apj, 647, 128

\bibitem[{{Evans} {et~al.}(2009){Evans}, {Dunham}, {J{\o}rgensen}, {Enoch},
  {Mer{\'{\i}}n}, {van Dishoeck}, {Alcal{\'a}}, {Myers}, {Stapelfeldt},
  {Huard}, {Allen}, {Harvey}, {van Kempen}, {Blake}, {Koerner}, {Mundy},
  {Padgett}, \& {Sargent}}]{eva09}
{Evans}, II, N.~J., {Dunham}, M.~M., {J{\o}rgensen}, J.~K., {Enoch}, M.~L.,
  {Mer{\'{\i}}n}, B., {van Dishoeck}, E.~F., {Alcal{\'a}}, J.~M., {Myers},
  P.~C., {Stapelfeldt}, K.~R., {Huard}, T.~L., {Allen}, L.~E., {Harvey}, P.~M.,
  {van Kempen}, T., {Blake}, G.~A., {Koerner}, D.~W., {Mundy}, L.~G.,
  {Padgett}, D.~L., \& {Sargent}, A.~I. 2009, \apjs, 181, 321

\bibitem[{{Feldmann} {et~al.}(2011){Feldmann}, {Gnedin}, \&
  {Kravtsov}}]{fel11b}
{Feldmann}, R., {Gnedin}, N.~Y., \& {Kravtsov}, A.~V. 2011, arXiv/1112.1732

\bibitem[{{F{\"o}rster Schreiber} {et~al.}(2009)}]{for09}
{F{\"o}rster Schreiber}, N.~M. {et~al.} 2009, \apj, 706, 1364

\bibitem[{{Fukui} \& {Kawamura}(2010)}]{fuk10}
{Fukui}, Y. \& {Kawamura}, A. 2010, \araa, 48, 547

\bibitem[{{Genzel} {et~al.}(2010)}]{gen10}
{Genzel}, R. {et~al.} 2010, \mnras, 407, 2091

\bibitem[{{Genzel} {et~al.}(2011{\natexlab{a}})}]{gen11b}
---. 2011{\natexlab{a}}, ArXiv e-prints

\bibitem[{{Genzel} {et~al.}(2011{\natexlab{b}})}]{gen11a}
---. 2011{\natexlab{b}}, \apj, 733, 101

\bibitem[{{Glover} {et~al.}(2010){Glover}, {Federrath}, {Mac Low}, \&
  {Klessen}}]{glo10}
{Glover}, S.~C.~O., {Federrath}, C., {Mac Low}, M., \& {Klessen}, R.~S. 2010,
  \mnras, 404, 2

\bibitem[{{Glover} \& {Mac Low}(2011)}]{glo11}
{Glover}, S.~C.~O. \& {Mac Low}, M.-M. 2011, \mnras, 412, 337

\bibitem[{{Goldsmith}(2001)}]{gol01}
{Goldsmith}, P.~F. 2001, \apj, 557, 736

\bibitem[{{Greve} {et~al.}(2005)}]{gre05}
{Greve}, T.~R. {et~al.} 2005, \mnras, 359, 1165

\bibitem[{{Groves} {et~al.}(2008){Groves}, {Dopita}, {Sutherland}, {Kewley},
  {Fischera}, {Leitherer}, {Brandl}, \& {van Breugel}}]{gro08}
{Groves}, B., {Dopita}, M.~A., {Sutherland}, R.~S., {Kewley}, L.~J.,
  {Fischera}, J., {Leitherer}, C., {Brandl}, B., \& {van Breugel}, W. 2008,
  \apjs, 176, 438

\bibitem[{{Guelin} {et~al.}(1993){Guelin}, {Zylka}, {Mezger}, {Haslam},
  {Kreysa}, {Lemke}, \& {Sievers}}]{gue93}
{Guelin}, M., {Zylka}, R., {Mezger}, P.~G., {Haslam}, C.~G.~T., {Kreysa}, E.,
  {Lemke}, R., \& {Sievers}, A.~W. 1993, \aap, 279, L37

\bibitem[{{Hailey-Dunsheath} {et~al.}(2008){Hailey-Dunsheath}, {Nikola},
  {Oberst}, {Parshley}, {Stacey}, {Farrah}, {Benford}, \& {Staguhn}}]{hai08}
{Hailey-Dunsheath}, S., {Nikola}, T., {Oberst}, T., {Parshley}, S., {Stacey},
  G.~J., {Farrah}, D., {Benford}, D.~J., \& {Staguhn}, J. 2008, in EAS
  Publications Series, Vol.~31, EAS Publications Series, ed. C.~{Kramer},
  S.~{Aalto}, \& R.~{Simon}, 159--162

\bibitem[{{Hayward} {et~al.}(2011){Hayward}, {Kere{\v s}}, {Jonsson},
  {Narayanan}, {Cox}, \& {Hernquist}}]{hay11}
{Hayward}, C.~C., {Kere{\v s}}, D., {Jonsson}, P., {Narayanan}, D., {Cox},
  T.~J., \& {Hernquist}, L. 2011, arXiv/1101.0002

\bibitem[{{Hayward} {et~al.}(2010){Hayward}, {Narayanan}, {Jonsson}, {Cox},
  {Kere{\v s}}, {Hopkins}, \& {Hernquist}}]{hay10}
{Hayward}, C.~C., {Narayanan}, D., {Jonsson}, P., {Cox}, T.~J., {Kere{\v s}},
  D., {Hopkins}, P.~F., \& {Hernquist}, L. 2010, Conference Proceedings for
  UP2010: Have Observations Revealed a Variable Upper End of the Initial Mass
  Function? Treyer, Lee, Seibert, Wyder, Neil eds. arXiv/1008.4584

\bibitem[{{Hernquist}(1989)}]{her89}
{Hernquist}, L. 1989, \nat, 340, 687

\bibitem[{{Hernquist}(1990)}]{her90}
---. 1990, \apj, 356, 359

\bibitem[{{Hinz} \& {Rieke}(2006)}]{hin06}
{Hinz}, J.~L. \& {Rieke}, G.~H. 2006, \apj, 646, 872

\bibitem[{{Hollenbach} \& {Tielens}(1999)}]{hol99}
{Hollenbach}, D.~J. \& {Tielens}, A.~G.~G.~M. 1999, Reviews of Modern Physics,
  71, 173

\bibitem[{{Hopkins} {et~al.}(2008{\natexlab{a}}){Hopkins}, {Cox}, \&
  {Hernquist}}]{hop08e}
{Hopkins}, P.~F., {Cox}, T.~J., \& {Hernquist}, L. 2008{\natexlab{a}}, \apj,
  689, 17

\bibitem[{{Hopkins} {et~al.}(2007{\natexlab{a}}){Hopkins}, {Hernquist}, {Cox},
  {Robertson}, \& {Krause}}]{hop07d}
{Hopkins}, P.~F., {Hernquist}, L., {Cox}, T.~J., {Robertson}, B., \& {Krause},
  E. 2007{\natexlab{a}}, \apj, 669, 67

\bibitem[{{Hopkins} {et~al.}(2007{\natexlab{b}}){Hopkins}, {Richards}, \&
  {Hernquist}}]{hop07}
{Hopkins}, P.~F., {Richards}, G.~T., \& {Hernquist}, L. 2007{\natexlab{b}},
  \apj, 654, 731

\bibitem[{{Hopkins} {et~al.}(2010){Hopkins}, {Younger}, {Hayward}, {Narayanan},
  \& {Hernquist}}]{hop10}
{Hopkins}, P.~F., {Younger}, J.~D., {Hayward}, C.~C., {Narayanan}, D., \&
  {Hernquist}, L. 2010, \mnras, 402, 1693

\bibitem[{{Hopkins} {et~al.}(2006{\natexlab{a}})}]{hop06}
{Hopkins}, P.~F. {et~al.} 2006{\natexlab{a}}, \apjs, 163, 1

\bibitem[{{Hopkins} {et~al.}(2006{\natexlab{b}})}]{hop06b}
---. 2006{\natexlab{b}}, \apjs, 163, 50

\bibitem[{{Hopkins} {et~al.}(2008{\natexlab{b}})}]{hop08a}
---. 2008{\natexlab{b}}, \apjs, 175, 356

\bibitem[{{Hopkins} {et~al.}(2008{\natexlab{c}})}]{hop08b}
---. 2008{\natexlab{c}}, \apjs, 175, 390

\bibitem[{{Hopkins} {et~al.}(2008{\natexlab{d}})}]{hop08c}
---. 2008{\natexlab{d}}, \apj, 679, 156

\bibitem[{{Hopkins} {et~al.}(2009)}]{hop09}
---. 2009, ArXiv e-prints

\bibitem[{{Hunter} {et~al.}(1997){Hunter}, {Bertsch}, {Catelli}, {Dame},
  {Digel}, {Dingus}, {Esposito}, {Fichtel}, {Hartman}, {Kanbach}, {Kniffen},
  {Lin}, {Mayer-Hasselwander}, {Michelson}, {von Montigny}, {Mukherjee},
  {Nolan}, {Schneid}, {Sreekumar}, {Thaddeus}, \& {Thompson}}]{hun97}
{Hunter}, S.~D., {Bertsch}, D.~L., {Catelli}, J.~R., {Dame}, T.~M., {Digel},
  S.~W., {Dingus}, B.~L., {Esposito}, J.~A., {Fichtel}, C.~E., {Hartman},
  R.~C., {Kanbach}, G., {Kniffen}, D.~A., {Lin}, Y.~C., {Mayer-Hasselwander},
  H.~A., {Michelson}, P.~F., {von Montigny}, C., {Mukherjee}, R., {Nolan},
  P.~L., {Schneid}, E., {Sreekumar}, P., {Thaddeus}, P., \& {Thompson}, D.~J.
  1997, \apj, 481, 205

\bibitem[{{Israel}(1997)}]{isr97}
{Israel}, F.~P. 1997, \aap, 328, 471

\bibitem[{{Jonsson}(2006)}]{jon06a}
{Jonsson}, P. 2006, \mnras, 372, 2

\bibitem[{{Jonsson} {et~al.}(2006){Jonsson}, {Cox}, {Primack}, \&
  {Somerville}}]{jon06b}
{Jonsson}, P., {Cox}, T.~J., {Primack}, J.~R., \& {Somerville}, R.~S. 2006,
  \apj, 637, 255

\bibitem[{{Jonsson} {et~al.}(2010){Jonsson}, {Groves}, \& {Cox}}]{jon10a}
{Jonsson}, P., {Groves}, B.~A., \& {Cox}, T.~J. 2010, \mnras, 186

\bibitem[{{Jonsson} \& {Primack}(2010)}]{jon10b}
{Jonsson}, P. \& {Primack}, J.~R. 2010, New Astronomy, 15, 509

\bibitem[{{Joung} {et~al.}(2009){Joung}, {Mac Low}, \& {Bryan}}]{jou09}
{Joung}, M.~R., {Mac Low}, M., \& {Bryan}, G.~L. 2009, \apj, 704, 137

\bibitem[{{Juvela} \& {Ysard}(2011)}]{juv11}
{Juvela}, M. \& {Ysard}, N. 2011, ArXiv e-prints

\bibitem[{{Kennicutt}(1998{\natexlab{a}})}]{ken98a}
{Kennicutt}, Jr., R.~C. 1998{\natexlab{a}}, \araa, 36, 189

\bibitem[{{Kennicutt}(1998{\natexlab{b}})}]{ken98b}
---. 1998{\natexlab{b}}, \apj, 498, 541

\bibitem[{{Kennicutt} {et~al.}(2007){Kennicutt}, {Calzetti}, {Walter}, {Helou},
  {Hollenbach}, {Armus}, {Bendo}, {Dale}, {Draine}, {Engelbracht}, \&
  {Gordon}}]{ken07}
{Kennicutt}, Jr., R.~C., {Calzetti}, D., {Walter}, F., {Helou}, G.,
  {Hollenbach}, D.~J., {Armus}, L., {Bendo}, G., {Dale}, D.~A., {Draine},
  B.~T., {Engelbracht}, C.~W., \& {Gordon}, K.~D. 2007, \apj, 671, 333

\bibitem[{{Kennicutt} {et~al.}(2003)}]{ken03}
{Kennicutt}, Jr., R.~C. {et~al.} 2003, \pasp, 115, 928

\bibitem[{{Keres} {et~al.}(2011){Keres}, {Vogelsberger}, {Sijacki}, {Springel},
  \& {Hernquist}}]{ker11}
{Keres}, D., {Vogelsberger}, M., {Sijacki}, D., {Springel}, V., \& {Hernquist},
  L. 2011, arXiv/1109.4638

\bibitem[{{Krumholz} \& {Dekel}(2010)}]{kru10}
{Krumholz}, M.~R. \& {Dekel}, A. 2010, \mnras, 406, 112

\bibitem[{{Krumholz} {et~al.}(2011{\natexlab{a}}){Krumholz}, {Dekel}, \&
  {McKee}}]{kru11c}
{Krumholz}, M.~R., {Dekel}, A., \& {McKee}, C.~F. 2011{\natexlab{a}},
  arXiv/1109.4150

\bibitem[{{Krumholz} \& {Gnedin}(2011)}]{kru11b}
{Krumholz}, M.~R. \& {Gnedin}, N.~Y. 2011, \apj, 729, 36

\bibitem[{{Krumholz} {et~al.}(2011{\natexlab{b}}){Krumholz}, {Leroy}, \&
  {McKee}}]{kru11a}
{Krumholz}, M.~R., {Leroy}, A.~K., \& {McKee}, C.~F. 2011{\natexlab{b}}, \apj,
  731, 25

\bibitem[{{Krumholz} \& {McKee}(2005)}]{kru05}
{Krumholz}, M.~R. \& {McKee}, C.~F. 2005, \apj, 630, 250

\bibitem[{{Krumholz} {et~al.}(2008){Krumholz}, {McKee}, \& {Tumlinson}}]{kru08}
{Krumholz}, M.~R., {McKee}, C.~F., \& {Tumlinson}, J. 2008, \apj, 689, 865

\bibitem[{{Krumholz} {et~al.}(2009{\natexlab{a}}){Krumholz}, {McKee}, \&
  {Tumlinson}}]{kru09a}
---. 2009{\natexlab{a}}, \apj, 693, 216

\bibitem[{{Krumholz} {et~al.}(2009{\natexlab{b}}){Krumholz}, {McKee}, \&
  {Tumlinson}}]{kru09b}
---. 2009{\natexlab{b}}, \apj, 699, 850

\bibitem[{{Krumholz} \& {Tan}(2007)}]{kru07b}
{Krumholz}, M.~R. \& {Tan}, J.~C. 2007, \apj, 654, 304

\bibitem[{{Krumholz} \& {Thompson}(2007)}]{kru07}
{Krumholz}, M.~R. \& {Thompson}, T.~A. 2007, \apj, 669, 289

\bibitem[{{Kutner} \& {Leung}(1985)}]{kut85}
{Kutner}, M.~L. \& {Leung}, C.~M. 1985, \apj, 291, 188

\bibitem[{{Larson}(1981)}]{lar81}
{Larson}, R.~B. 1981, \mnras, 194, 809

\bibitem[{{Lee} {et~al.}(1996){Lee}, {Bettens}, \& {Herbst}}]{lee96}
{Lee}, H., {Bettens}, R.~P.~A., \& {Herbst}, E. 1996, \aaps, 119, 111

\bibitem[{{Leitherer} {et~al.}(1999)}]{lei99}
{Leitherer}, C. {et~al.} 1999, \apjs, 123, 3

\bibitem[{{Lemaster} \& {Stone}(2008)}]{lem08}
{Lemaster}, M.~N. \& {Stone}, J.~M. 2008, \apjl, 682, L97

\bibitem[{{Leroy} {et~al.}(2006){Leroy}, {Bolatto}, {Walter}, \&
  {Blitz}}]{ler06}
{Leroy}, A., {Bolatto}, A., {Walter}, F., \& {Blitz}, L. 2006, \apj, 643, 825

\bibitem[{{Leroy} {et~al.}(2011)}]{ler11}
{Leroy}, A.~K. {et~al.} 2011, arXiv/1102.4618

\bibitem[{{Lombardi} {et~al.}(2006){Lombardi}, {Alves}, \& {Lada}}]{lom06}
{Lombardi}, M., {Alves}, J., \& {Lada}, C.~J. 2006, \aap, 454, 781

\bibitem[{{Magdis} {et~al.}(2011)}]{mag11}
{Magdis}, G.~E. {et~al.} 2011, arXiv/1109.1140

\bibitem[{{Maloney} \& {Black}(1988)}]{mal88}
{Maloney}, P. \& {Black}, J.~H. 1988, \apj, 325, 389

\bibitem[{{Markwardt}(2009)}]{mar09b}
{Markwardt}, C.~B. 2009, in Astronomical Society of the Pacific Conference
  Series, Vol. 411, Astronomical Data Analysis Software and Systems XVIII, ed.
  {D.~A.~Bohlender, D.~Durand, \& P.~Dowler}, 251--+

\bibitem[{{McKee} \& {Krumholz}(2010)}]{mck10}
{McKee}, C.~F. \& {Krumholz}, M.~R. 2010, \apj, 709, 308

\bibitem[{{McKee} \& {Ostriker}(1977)}]{mck77}
{McKee}, C.~F. \& {Ostriker}, J.~P. 1977, \apj, 218, 148

\bibitem[{{Meier} {et~al.}(2010){Meier}, {Turner}, {Beck}, {Gorjian}, {Tsai},
  \& {Van Dyk}}]{mei10}
{Meier}, D.~S., {Turner}, J.~L., {Beck}, S.~C., {Gorjian}, V., {Tsai}, C., \&
  {Van Dyk}, S.~D. 2010, \aj, 140, 1294

\bibitem[{{Meijerink} {et~al.}(2007){Meijerink}, {Spaans}, \& {Israel}}]{mei07}
{Meijerink}, R., {Spaans}, M., \& {Israel}, F.~P. 2007, \aap, 461, 793

\bibitem[{{Mihos} \& {Hernquist}(1994{\natexlab{a}})}]{mih94b}
{Mihos}, J.~C. \& {Hernquist}, L. 1994{\natexlab{a}}, \apj, 437, 611

\bibitem[{{Mihos} \& {Hernquist}(1994{\natexlab{b}})}]{mih94a}
---. 1994{\natexlab{b}}, \apjl, 431, L9

\bibitem[{{Mihos} \& {Hernquist}(1996)}]{mih96}
---. 1996, \apj, 464, 641

\bibitem[{{Mo} {et~al.}(1998){Mo}, {Mao}, \& {White}}]{mo98}
{Mo}, H.~J., {Mao}, S., \& {White}, S.~D.~M. 1998, \mnras, 295, 319

\bibitem[{{Moster} {et~al.}(2011{\natexlab{a}}){Moster}, {Macci{\`o}},
  {Somerville}, {Naab}, \& {Cox}}]{mos11a}
{Moster}, B.~P., {Macci{\`o}}, A.~V., {Somerville}, R.~S., {Naab}, T., \&
  {Cox}, T.~J. 2011{\natexlab{a}}, \mnras, 415, 3750

\bibitem[{{Moster} {et~al.}(2011{\natexlab{b}}){Moster}, {Maccio'},
  {Somerville}, {Naab}, \& {Cox}}]{mos11b}
{Moster}, B.~P., {Maccio'}, A.~V., {Somerville}, R.~S., {Naab}, T., \& {Cox},
  T.~J. 2011{\natexlab{b}}, arXiv/1108.1796

\bibitem[{{Narayanan} {et~al.}(2011{\natexlab{a}}){Narayanan}, {Cox},
  {Hayward}, \& {Hernquist}}]{nar11a}
{Narayanan}, D., {Cox}, T.~J., {Hayward}, C.~C., \& {Hernquist}, L.
  2011{\natexlab{a}}, \mnras, 412, 287

\bibitem[{{Narayanan} {et~al.}(2009){Narayanan}, {Cox}, {Hayward}, {Younger},
  \& {Hernquist}}]{nar09}
{Narayanan}, D., {Cox}, T.~J., {Hayward}, C.~C., {Younger}, J.~D., \&
  {Hernquist}, L. 2009, \mnras, 400, 1919

\bibitem[{{Narayanan} {et~al.}(2010{\natexlab{a}}){Narayanan}, {Dey},
  {Hayward}, {Cox}, {Bussmann}, {Brodwin}, {Jonsson}, {Hopkins}, {Groves},
  {Younger}, \& {Hernquist}}]{nar10b}
{Narayanan}, D., {Dey}, A., {Hayward}, C.~C., {Cox}, T.~J., {Bussmann}, R.~S.,
  {Brodwin}, M., {Jonsson}, P., {Hopkins}, P.~F., {Groves}, B., {Younger},
  J.~D., \& {Hernquist}, L. 2010{\natexlab{a}}, \mnras, 407, 1701

\bibitem[{{Narayanan} {et~al.}(2010{\natexlab{b}}){Narayanan}, {Hayward},
  {Cox}, {Hernquist}, {Jonsson}, {Younger}, \& {Groves}}]{nar10a}
{Narayanan}, D., {Hayward}, C.~C., {Cox}, T.~J., {Hernquist}, L., {Jonsson},
  P., {Younger}, J.~D., \& {Groves}, B. 2010{\natexlab{b}}, \mnras, 401, 1613

\bibitem[{{Narayanan} {et~al.}(2011{\natexlab{b}}){Narayanan}, {Krumholz},
  {Ostriker}, \& {Hernquist}}]{nar11b}
{Narayanan}, D., {Krumholz}, M., {Ostriker}, E.~C., \& {Hernquist}, L.
  2011{\natexlab{b}}, \mnras, 418, 664

\bibitem[{{Narayanan} {et~al.}(2006){Narayanan}, {Kulesa}, {Boss}, \&
  {Walker}}]{nar06b}
{Narayanan}, D., {Kulesa}, C.~A., {Boss}, A., \& {Walker}, C.~K. 2006, \apj,
  647, 1426

\bibitem[{{Obreschkow} \& {Rawlings}(2009)}]{obr09c}
{Obreschkow}, D. \& {Rawlings}, S. 2009, \mnras, 394, 1857

\bibitem[{{Oka} {et~al.}(1998){Oka}, {Hasegawa}, {Hayashi}, {Handa}, \&
  {Sakamoto}}]{oka98}
{Oka}, T., {Hasegawa}, T., {Hayashi}, M., {Handa}, T., \& {Sakamoto}, S. 1998,
  \apj, 493, 730

\bibitem[{{Ostriker} {et~al.}(2010){Ostriker}, {McKee}, \& {Leroy}}]{ost10}
{Ostriker}, E.~C., {McKee}, C.~F., \& {Leroy}, A.~K. 2010, \apj, 721, 975

\bibitem[{{Ostriker} \& {Shetty}(2011)}]{ost11}
{Ostriker}, E.~C. \& {Shetty}, R. 2011, \apj, 731, 41

\bibitem[{{Ostriker} {et~al.}(2001){Ostriker}, {Stone}, \& {Gammie}}]{ost01}
{Ostriker}, E.~C., {Stone}, J.~M., \& {Gammie}, C.~F. 2001, \apj, 546, 980

\bibitem[{{Padoan} \& {Nordlund}(2002)}]{pad02}
{Padoan}, P. \& {Nordlund}, {\AA}. 2002, \apj, 576, 870

\bibitem[{{Papadopoulos} {et~al.}(2011){Papadopoulos}, {Thi}, {Miniati}, \&
  {Viti}}]{pap10b}
{Papadopoulos}, P.~P., {Thi}, W.-F., {Miniati}, F., \& {Viti}, S. 2011, \mnras,
  414, 1705

\bibitem[{{Papadopoulos} {et~al.}(2002){Papadopoulos}, {Thi}, \&
  {Viti}}]{pap02b}
{Papadopoulos}, P.~P., {Thi}, W.-F., \& {Viti}, S. 2002, \apj, 579, 270

\bibitem[{{Pineda} {et~al.}(2008){Pineda}, {Caselli}, \& {Goodman}}]{pin08}
{Pineda}, J.~E., {Caselli}, P., \& {Goodman}, A.~A. 2008, \apj, 679, 481

\bibitem[{{Price} {et~al.}(2011){Price}, {Federrath}, \& {Brunt}}]{pri11}
{Price}, D.~J., {Federrath}, C., \& {Brunt}, C.~M. 2011, \apjl, 727, L21+

\bibitem[{{Robertson} {et~al.}(2004){Robertson}, {Yoshida}, {Springel}, \&
  {Hernquist}}]{rob04}
{Robertson}, B., {Yoshida}, N., {Springel}, V., \& {Hernquist}, L. 2004, \apj,
  606, 32

\bibitem[{{Robertson} {et~al.}(2006)}]{rob06a}
{Robertson}, B. {et~al.} 2006, \apj, 641, 21

\bibitem[{{Robitaille} \& {Whitney}(2010)}]{rob10}
{Robitaille}, T.~P. \& {Whitney}, B.~A. 2010, \apjl, 710, L11

\bibitem[{{Rosolowsky}(2005)}]{ros05}
{Rosolowsky}, E. 2005, \pasp, 117, 1403

\bibitem[{{Rosolowsky}(2007)}]{ros07}
---. 2007, \apj, 654, 240

\bibitem[{{Rosolowsky} {et~al.}(2003){Rosolowsky}, {Engargiola}, {Plambeck}, \&
  {Blitz}}]{ros03}
{Rosolowsky}, E., {Engargiola}, G., {Plambeck}, R., \& {Blitz}, L. 2003, \apj,
  599, 258

\bibitem[{{Sakamoto} {et~al.}(1999)}]{sak99}
{Sakamoto}, K. {et~al.} 1999, \apj, 514, 68

\bibitem[{{Schaye} \& {Dalla Vecchia}(2008)}]{sch08}
{Schaye}, J. \& {Dalla Vecchia}, C. 2008, \mnras, 383, 1210

\bibitem[{{Schmidt}(1959)}]{sch59}
{Schmidt}, M. 1959, \apj, 129, 243

\bibitem[{{Schuster} {et~al.}(2007){Schuster}, {Kramer}, {Hitschfeld},
  {Garcia-Burillo}, \& {Mookerjea}}]{sch07}
{Schuster}, K.~F., {Kramer}, C., {Hitschfeld}, M., {Garcia-Burillo}, S., \&
  {Mookerjea}, B. 2007, \aap, 461, 143

\bibitem[{{Scoville} {et~al.}(1991){Scoville}, {Sargent}, {Sanders}, \&
  {Soifer}}]{sco91}
{Scoville}, N.~Z., {Sargent}, A.~I., {Sanders}, D.~B., \& {Soifer}, B.~T. 1991,
  \apjl, 366, L5

\bibitem[{{Shapley}(2011)}]{sha11}
{Shapley}, A.~E. 2011, ARA\&A in press; arXiv/1107.5060

\bibitem[{{Shapley} {et~al.}(2004){Shapley}, {Erb}, {Pettini}, {Steidel}, \&
  {Adelberger}}]{sha04}
{Shapley}, A.~E., {Erb}, D.~K., {Pettini}, M., {Steidel}, C.~C., \&
  {Adelberger}, K.~L. 2004, \apj, 612, 108

\bibitem[{{Shetty} {et~al.}(2011{\natexlab{a}})}]{she11b}
{Shetty}, R. {et~al.} 2011{\natexlab{a}}, arXiv/1104.3695

\bibitem[{{Shetty} {et~al.}(2011{\natexlab{b}})}]{she11a}
---. 2011{\natexlab{b}}, \mnras, 412, 1686

\bibitem[{{Sijacki} {et~al.}(2011){Sijacki}, {Vogelsberger}, {Keres},
  {Springel}, \& {Hernquist}}]{sij11}
{Sijacki}, D., {Vogelsberger}, M., {Keres}, D., {Springel}, V., \& {Hernquist},
  L. 2011, arXiv/1109.3468

\bibitem[{{Snyder} {et~al.}(2011){Snyder}, {Cox}, {Hayward}, {Hernquist}, \&
  {Jonsson}}]{sny11}
{Snyder}, G.~F., {Cox}, T.~J., {Hayward}, C.~C., {Hernquist}, L., \& {Jonsson},
  P. 2011, arXiv/1102.3689

\bibitem[{{Solomon} {et~al.}(1997){Solomon}, {Downes}, {Radford}, \&
  {Barrett}}]{sol97}
{Solomon}, P.~M., {Downes}, D., {Radford}, S.~J.~E., \& {Barrett}, J.~W. 1997,
  \apj, 478, 144

\bibitem[{{Solomon} {et~al.}(1987){Solomon}, {Rivolo}, {Barrett}, \&
  {Yahil}}]{sol87}
{Solomon}, P.~M., {Rivolo}, A.~R., {Barrett}, J., \& {Yahil}, A. 1987, \apj,
  319, 730

\bibitem[{{Springel}(2000)}]{spr00}
{Springel}, V. 2000, \mnras, 312, 859

\bibitem[{{Springel}(2005)}]{spr05b}
---. 2005, \mnras, 364, 1105

\bibitem[{{Springel}(2010)}]{spr10}
---. 2010, \mnras, 401, 791

\bibitem[{{Springel} {et~al.}(2005){Springel}, {Di Matteo}, \&
  {Hernquist}}]{spr05a}
{Springel}, V., {Di Matteo}, T., \& {Hernquist}, L. 2005, \mnras, 361, 776

\bibitem[{{Springel} \& {Hernquist}(2003)}]{spr03a}
{Springel}, V. \& {Hernquist}, L. 2003, \mnras, 339, 289

\bibitem[{{Sternberg} \& {Dalgarno}(1995)}]{ste95}
{Sternberg}, A. \& {Dalgarno}, A. 1995, \apjs, 99, 565

\bibitem[{{Strong} \& {Mattox}(1996)}]{str96}
{Strong}, A.~W. \& {Mattox}, J.~R. 1996, \aap, 308, L21

\bibitem[{{Tacconi} {et~al.}(2006)}]{tac06}
{Tacconi}, L.~J. {et~al.} 2006, \apj, 640, 228

\bibitem[{{Tacconi} {et~al.}(2008)}]{tac08}
---. 2008, \apj, 680, 246

\bibitem[{{Tacconi} {et~al.}(2010)}]{tac10}
---. 2010, \nat, 463, 781

\bibitem[{{Tan}(2010)}]{tan10}
{Tan}, J.~C. 2010, \apjl, 710, L88

\bibitem[{{Torrey} {et~al.}(2011){Torrey}, {Cox}, {Kewley}, \&
  {Hernquist}}]{tor11}
{Torrey}, P., {Cox}, T.~J., {Kewley}, L., \& {Hernquist}, L. 2011,
  arXiv/1107.0001

\bibitem[{{V{\'a}zquez} \& {Leitherer}(2005)}]{vaz05}
{V{\'a}zquez}, G.~A. \& {Leitherer}, C. 2005, \apj, 621, 695

\bibitem[{{Vladilo}(1998)}]{vla98}
{Vladilo}, G. 1998, \apj, 493, 583

\bibitem[{{Vogelsberger} {et~al.}(2011){Vogelsberger}, {Sijacki}, {Keres},
  {Springel}, \& {Hernquist}}]{vog11}
{Vogelsberger}, M., {Sijacki}, D., {Keres}, D., {Springel}, V., \& {Hernquist},
  L. 2011, arXiv/1109.1281

\bibitem[{{Wall}(2007)}]{wal07}
{Wall}, W.~F. 2007, \mnras, 379, 674

\bibitem[{{Watson}(2011)}]{wat11}
{Watson}, D. 2011, arXiv/1107.6031

\bibitem[{{Weingartner} \& {Draine}(2001)}]{wei01}
{Weingartner}, J.~C. \& {Draine}, B.~T. 2001, \apj, 548, 296

\bibitem[{{Wilson}(1995)}]{wil95}
{Wilson}, C.~D. 1995, \apjl, 448, L97+

\bibitem[{{Wolfire} {et~al.}(2010){Wolfire}, {Hollenbach}, \& {McKee}}]{wol10}
{Wolfire}, M.~G., {Hollenbach}, D., \& {McKee}, C.~F. 2010, \apj, 716, 1191

\bibitem[{{Wong} \& {Blitz}(2002)}]{won02}
{Wong}, T. \& {Blitz}, L. 2002, \apj, 569, 157

\bibitem[{{Younger} {et~al.}(2009){Younger}, {Hayward}, {Narayanan}, {Cox},
  {Hernquist}, \& {Jonsson}}]{you09}
{Younger}, J.~D., {Hayward}, C.~C., {Narayanan}, D., {Cox}, T.~J., {Hernquist},
  L., \& {Jonsson}, P. 2009, \mnras, 396, L66

\end{thebibliography}
\end{document}